\documentclass{aastex}
\usepackage{emulateapj5,epsfig}

\def\be{\begin{equation}}
\def\ee{\end{equation}}
\def\ba{\begin{eqnarray}}
\def\ea{\end{eqnarray}}
\def\go{\mathrel{\raise.3ex\hbox{$>$}\mkern-14mu
             \lower0.6ex\hbox{$\sim$}}}
\def\lo{\mathrel{\raise.3ex\hbox{$<$}\mkern-14mu
             \lower0.6ex\hbox{$\sim$}}}

\def\beps{{\mbox{\boldmath $\epsilon$}}}

\def\bmu{{\mbox{\boldmath $\mu$}}}
\def\bOmega{{\mbox{\boldmath $\Omega$}}}
\def\bE{{\bf E}}
\def\bB{{\bf B}}
\def\bA{{\bf A}}
\def\bS{{\bf S}}
\def\bs{{\bf s}}
\def\br{{\bf r}}
\def\bk{{\bf k}}

\def\bT{{\bf T}}

\def\bI{{\bf I}}
\def\hatk{{\hat {\bf k}}}
\def\hate{{\hat {\bf e}}}
\def\cA{{\cal A}}
\def\bcI{{\mbox{\boldmath ${\cal I}$}}}
\def\bcM{{\mbox{\boldmath ${\cal M}$}}}
\def\bcS{{\mbox{\boldmath ${\cal S}$}}}
\def\xp{{x'}}
\def\yp{{y'}}

\def\Cdp{{C^{(e)}({\Delta\phi})}}
\def\Sdp{{S^{(e)}({\Delta\phi})}}
\def\Cidp{{C^{(i)}({\Delta\phi})}}
\def\Sidp{{S^{(i)}({\Delta\phi})}}

\newcommand{\lp}{\left(}
\newcommand{\rp}{\right)}
\newcommand{\lb}{\left[}
\newcommand{\rb}{\right]}
\newcommand{\etal}{et al.}
\newcommand{\thetab}{\theta_B}

\newcommand{\polarb}{\beta}
\newcommand{\uel}{u_{\rm e}}
\newcommand{\uion}{u_{\rm i}}
\newcommand{\vel}{v_{\rm e}}
\newcommand{\vpa}{a}
\newcommand{\vpq}{q}
\newcommand{\vpm}{m}
\newcommand{\Evp}{E_{\rm V}}

\newcommand{\polarbvp}{\beta_{\rm V}}
\newcommand{\gamdampe}{\gamma_{\rm e}}
\newcommand{\Emcp}{E_{\rm Far}}
\newcommand{\thetamcp}{\theta_{\rm coll}}

\shorttitle{Transfer of Polarized Radiation}
\begin{document}

\title{Transfer of Polarized Radiation in 
Strongly Magnetized Plasmas and Thermal Emission from Magnetars: 
Effect of Vacuum Polarization}
\author{Dong Lai and Wynn C.~G. Ho}
\affil{Center for Radiophysics and Space Research,
Department of Astronomy, Cornell University, Ithaca, NY 14853\\
E-mail: dong, wynnho@astro.cornell.edu}

\begin{abstract}
We present a theoretical study of radiative transfer in strongly magnetized
electron-ion plasmas, focusing on the effect of vacuum polarization
due to quantum electrodynamics. This study is directly relevant to thermal
radiation from the surfaces of highly magnetized neutron stars, which have 
been detected in recent years. 
Strong-field vacuum polarization modifies the photon propagation modes 
in the plasma, and induces a ``vacuum resonance'' at
which a polarized X-ray photon propagating outward in the neutron star
atmosphere can convert from a low-opacity mode to a high-opacity mode 
and vice versa. The effectiveness of this mode conversion depends on the 
photon energy and the atmosphere density gradient. For a wide range of 
field strengths, $7\times 10^{13}\lo B\lo {\rm a~few}\times 10^{16}$~G, 
the vacuum resonance lies between the photospheres of the two photon modes, 
and the emergent radiation spectrum from the neutron star is significantly 
modified by the vacuum resonance. (For lower field strengths, only the 
polarization spectrum is affected.) Under certain conditions, which depend 
on the field strength, photon energy and propagation direction, the vacuum 
resonance is accompanied by the phenomenon of mode collapse (at which the 
two photon modes become degenerate) and the breakdown of Faraday 
depolarization. Thus, the widely used description
of radiative transfer based on photon modes is not adequate to treat 
the vacuum polarization effect rigorously. We study the evolution 
of polarized X-rays across the vacuum resonance and derive
the transfer equation for the photon intensity matrix (Stokes parameters), 
taking into account the effect of birefringence of the plasma-vacuum medium, 
free-free absorption, and scatterings by electrons and ions. 
\end{abstract}
\keywords{magnetic fields -- radiative transfer -- stars: neutron
-- stars: atmospheres -- X-rays: stars}

\section{Introduction}

It has long been known that in a strong magnetic field
even the vacuum has a nontrivial dielectric property;
this is the result of vacuum polarization, a genuine quantum electrodynamics
effect (Heisenberg \& Euler 1936; Schwinger 1951; Adler 1971; 
Tsai \& Erber 1975; Heyl \& Hernquist 1997). This effect is usually negligible
for $B\lo B_Q$, but becomes increasingly important when $B$ exceeds $B_Q$, 
where $B_Q=m_e^2c^3/(e\hbar)=4.414\times 10^{13}$~G is the magnetic field at
which the electron cyclotron energy $\hbar\omega_{Be}=\hbar eB/(m_ec)$ equals
$m_ec^2$. A number of studies in the late 1970s showed that vacuum 
polarization can affect radiative transfer in magnetized plasmas by modifying
photon polarization modes and inducing features in the radiative opacities 
(Gnedin, Pavlov \& Shibanov 1978a,b; Meszaros \& Ventura 1978,1979; Pavlov \&
Shibanov 1979; Ventura, Nagel \& Meszaros 1979; see Pavlov \& Gnedin 1984 and
Meszaros 1992 for review). These earlier studies were mostly applied to radio
pulsars and accreting X-ray binaries, for which $B\lo B_Q$. Not surprisingly,
the effect of vacuum polarization in these systems appears to be relatively
minor and, as far as we know, no concrete observational manifestation of 
vacuum polarization has been identified. 

In the last few years, soft gamma-ray repeaters (SGRs) and anomalous X-ray
pulsars (AXPs) have been identified as a potentially new class of neutron stars
(NSs) endowed with superstrong ($B\go 10^{14}$~G) 
magnetic fields, the so-called ``magnetars'' (see Thompson \& Duncan 1996;
Hurley 2000; Thompson 2001; Israel et al 2001). This has stimulated new 
studies on radiative processes in the superstrong magnetic field regime.
Of particular relevance to the present paper is the detection of 
{\it thermal} X-rays from the surface of isolated magnetic NSs
(see Pavlov et al.~2002 for a recent review), 
including five magnetar candidates (4 AXPs and one SGR; see, e.g., 
Patel et al.~2001; 
Juett et al.~2002; Tiengo et al.~2002). The thermal emission can potentially
provide useful constraints on the interior physics, surface magnetic field
and composition of NSs. Several recent works have begun to model the
atmosphere emission from magnetars, including the effect of vacuum polarization
(Bezchastnov et al.~1996; Bulik \& Miller 1997; \"Ozel
2001; Ho \& Lai 2001,~2003; Lai \& Ho 2002a; a review/comment
of these works can be found in Ho \& Lai 2003).

In a highly magnetized plasma which characterizes NS atmospheres, 
there are two photon polarization modes, and they have very different
radiative opacities. Previous theoretical modeling of magnetic NS
atmospheres (see Ho \& Lai 2003 and references therein; see
also Pavlov et al.~1995; Zavlin \& Pavlov 2002 for reviews) 
are based on solving the transfer
equations for the two photon modes. As we discuss and quantify in 
\S 2 and \S3 below, this modal description of the radiative transport 
is, in general, inadequate to handle the vacuum polarization effect near the
``vacuum resonance'' where the modes can mix and even ``collapse''. 
The main purpose of this paper is to clarify these
issues and to derive general transfer equations that can be used
to study the transport of polarized radiation in the superstrong field regime. 

Section 2 contains a physical discussion of the vacuum polarization
effect on radiative transfer in magnetized NS atmospheres. 
In \S3, we derive the basic properties of photon modes in
magnetized plasmas and quantify the limitation of the modal
description of radiative transport. We then study in \S 4 and \S 5 the 
propagation of polarized photons in an inhomogeneous plasma.
We derive in \S6 the transfer equations for the photon intensity
matrix and conclude in \S 7.

\section{Vacuum Polarization Effect on Radiative Transfer and Atmosphere
Spectra of Magnetars: A Physical Discussion}

To motivate our quantitative study in \S 3-\S6, 
we present a physical discussion of the effect of vacuum polarization on 
radiative transfer in strong magnetic fields and on the thermal 
spectra of magnetars (see Lai \& Ho 2002a; Ho \& Lai 2003; hereafter 
LH02 and HL03). This also serves to clarify some of the confusions
that have appeared recently (see Lai \& Ho 2002b). 

In a highly magnetized NS atmosphere, both the plasma and vacuum polarization
contribute to the dielectric property of the medium. A ``vacuum
resonance'' arises when these two contributions ``compensate'' each other
(Gnedin et al.~1978a,b; Pavlov \& Shibanov 1979; Meszaros \& Ventura 1979; 
Ventura et al.~1979). For a photon of energy $E$, the vacuum 
resonance occurs at the density 
\be
\rho_V\simeq 0.964\,Y_e^{-1}B_{14}^2E_1^2 f^{-2}~{\rm g~cm}^{-3},
\label{eq:densvp}
\ee
where $Y_e$ is the electron fraction, $E_1=E/(1~{\rm keV})$,
$B=10^{14}B_{14}$~G is the magnetic field
strength, and $f=f(B)$ is a 
slowly varying function of $B$ and is of order unity
(LH02 and HL03). 
For $\rho>\rho_V$ (where the plasma effect dominates the dielectric
tensor) and $\rho<\rho_V$ (where vacuum polarization dominates), the 
photon modes (for $E$ much smaller than the electron cyclotron energy
$E_{Be}$) are almost linearly polarized (see Fig.~3 below): the extraordinary 
mode (X-mode) has its electric field vector $\bE$
perpendicular to the $\hatk$-${\hat\bB}$ plane, while the ordinary mode 
(O-mode) is polarized along the $\hatk$-$\hat\bB$ plane
(where $\hatk$ specifies the direction of photon propagation, $\hat\bB$ 
is the unit vector along the magnetic field). Near $\rho=\rho_V$, however,
the normal modes become circularly polarized as a result of the
``cancellation'' of the plasma and vacuum effects --- both effects tend to 
make the mode linearly polarized, but in mutually orthogonal directions. 
When a photon propagates in an inhomogeneous medium, its polarization
state will evolve adiabatically if the density variation is
sufficiently gentle. Thus, a X-mode (O-mode) photon will be converted
into a O-mode (X-mode) as it traverses the vacuum resonance (see Fig.~1).
This resonant mode conversion is analogous to the MSW effect of neutrino
oscillation (e.g., Haxton 1995). For this conversion to be effective, 
the adiabatic condition must be satisfied:
\ba
&&E\go E_{\rm ad}(B,\theta_B,H_\rho)\nonumber\\
&&\qquad =1.49\,\bigl(f\,\tan\theta_B |1-u_i|\bigr)^{2/3}
\left({5\,{\rm cm}\over H_\rho}\right)^{1/3}~{\rm keV},
\label{condition}\ea
where $\theta_B$ is the angle between ${\hatk}$ and ${\hat\bB}$, $u_i
=(E_{Bi}/E)^2$, $E_{Bi}$ is the ion cyclotron energy,
and $H_\rho=|dz/d\ln\rho|$ is the density scale
height (evaluated at $\rho=\rho_V$) along the ray.
For an ionized Hydrogen atmosphere, $H_\rho\simeq 2kT/(m_pg\cos\theta)
=1.65\,T_6/(g_{14}\cos\theta)$~cm, where $T=10^6\,T_6$~K is the temperature,
$g=10^{14}g_{14}$~cm~s$^{-2}$ is the gravitational acceleration, and $\theta$
is the angle between the ray and the surface normal.
We note that in this paper [as in our previous papers (LH02, HL03), and in 
several other references such as Meszaros \& Ventura (1979), 
Meszaros (1992, p.~94), etc.],
we refer to the O-mode as the mode with $|K|=|E_x/E_y|\gg 1$ and 
the X-mode as the mode with $|K|\ll 1$ (here the $z$-axis is along $\hatk$ and
the $y$-axis is in the direction of $\hat\bB\times \hatk$), thus
the name ``mode conversion''. Alternatively, one can define modes with 
definite helicity (the sign of $K=-iE_x/E_y$) such that $K$ changes
continuously as $\rho$ changes; we call these plus-mode and 
minus-mode (see \S 3). Thus we may also say that in the adiabatic limit, the
photon will remain in the same plus or minus branch, but the character of
the mode is changed across the vacuum resonance. Indeed, in the 
literature on radio wave propagation in plasmas (e.g., Budden 1961;
Zheleznyakov et al.~1983), the nonadiabatic case, in which the photon state
jumps across the continuous curves, is referred to as ``linear mode coupling''.
It is important to note that the ``mode conversion'' effect discussed here
is not a matter of semantics. The key point is that in the adiabatic limit, 
the photon polarization ellipse changes its orientation across the vacuum
resonance, and therefore the photon opacity changes significantly
\footnote{The O-mode has a significant component of its $\bE$ field along 
$\hat\bB$ (for most directions of propagation except when $\hatk$ is 
nearly parallel to $\hat\bB$), and therefore the O-mode
opacity is close to the $B=0$ value, while the X-mode opacity 
is much smaller.}. We prefer to use the term ``mode conversion'' since it
captures this essential feature important for radiative transfer, and it is
analogous to the MSW effect and other quantum mechanics problems involving
adiabatic evolution of quantum states (Landau 1932; Zener 1932).

Because the two photon modes have vastly different opacities,
the vacuum polarization-induced mode conversion can significantly affect
radiative transfer in magnetar atmospheres. The main effect of
vacuum polarization on the atmosphere spectrum can be understood
as follows. When the vacuum polarization effect is neglected, 
the decoupling densities of the O-mode and X-mode photons 
(i.e., the densities of their respective photospheres)
are approximately given by (for Hydrogen plasma) 
$\rho_O\simeq 0.42\,T_6^{-1/4}E_1^{3/2}G^{-1/2}$~g~cm$^{-3}$ and
$\rho_X\simeq 486\,T_6^{-1/4}E_1^{1/2}B_{14}G^{-1/2}$~g~cm$^{-3}$ (see LH02), 
where $G=1-e^{-E/kT}$. The vacuum resonance lies between these two photospheres
when $\rho_O<\rho_V<\rho_X$, i.e.
\be
B_l<B<B_h,
\ee
where
\ba
&&B_l\simeq 6.6\times 10^{13}\,f\,
T_6^{-1/8}E_1^{-1/4}G^{-1/4}~{\rm G},\\
&&B_h\simeq 5.1\times 10^{16}
\,f^2\,T_6^{-1/4}E_1^{-3/2}G^{-1/2}~{\rm G}.
\ea
When this condition is satisfied, the effective decoupling depths of
the photons are changed\footnote{For $B>B_h$ the vacuum resonance lies
deeper than the photospheres of both modes, while for $B<B_l$
the resonance lies outside both photospheres. In both cases,
the effect of vacuum polarization on the radiative spectrum is expected
to be small, although in the latter case ($B<B_l$), the polarization
of the emitted photons will be modified by the vacuum resonance.}.
Indeed, we see from Fig.~2 that mode conversion
makes the effective decoupling density of X-mode photons (which carry the
bulk of the thermal energy) smaller, thereby depleting the high-energy
tail of the spectrum and making the spectrum closer to black-body (although
the spectrum is still harder than black-body because of nongrey opacities)
\footnote{Even when mode conversion is neglected, the X-mode
decoupling depth can still be affected by vacuum polarization.
This is because the X-mode opacity exhibits a spike feature 
near the resonance, and the optical depth across the resonance region
can be significant;  see LH02.}.
This expectation is borne out in self-consistent atmosphere modeling
presented in HL03. Another important effect of vacuum polarization on the
spectrum, first noted in HL03, is the suppression of proton cyclotron lines
(and maybe other spectral lines; see also Ho et al.~2003). 
The physical origin for such line
suppression is related to the depletion of continuum flux, which makes
the decoupling depths inside and outside the line similar. HL03 suggests
that the absence of lines in the observed spectra of several AXPs may be an
indication of the vacuum polarization effect at work in these systems.

Our previous study (LH02) on the vacuum-induced mode conversion 
did not explicitly take into account of the effect of dissipation 
on the mode structure. Although this dissipative effect
is small under many situations, near the
vacuum resonance and for some directions of propagation,
the two photon modes can ``collapse'', i.e., they become identical and 
hence nonorthogonal (Soffel et al.~1983; see also Pavlov \& Shibanov 1979). 
The analysis of LH02 breaks down near these ``mode collapse'' points
(see \S3). The question therefore arises as to how the polarization state
evolves as the photon travels near the mode collapse point.
Also, as mentioned in \S 1, previous studies of NS atmospheres
with strong magnetic fields rely on the modal 
description of the radiative transport. This is valid
only in the limit of large Faraday depolarization (Gnedin \& Pavlov 1974),
which is not always satisfied near the vacuum resonance 
especially for superstrong field strengths (see \S 3). 
More importantly, the transfer equations based on normal modes
cannot handle the cases in which partial mode coupling (conversion)
occurs across
the vacuum resonance (i.e., when the adiabatic condition is neither strongly
satisfied or violated). Clearly, to properly account 
for the effects of mode collapse, breakdown
of Faraday depolarization, and mode conversion
associated with vacuum polarization, one must go beyond the 
modal description of the radiation field by formulating 
and solving the transfer equation in terms of
the photon intensity matrix (or Stokes parameters)
and including the birefringence of the plasma-vacuum medium 
(see \S 5 and \S6).

\section{Photon Polarization Modes, Vacuum Resonance,
Mode Collapse, and Faraday Depolarization}

The property of photon modes in a magnetized electron-ion plasma
including vacuum polarization was studied in LH02 and HL03
(see also Meszaros 1992 and references therein for earlier works in which 
electron plasmas with $B\ll B_Q$ were studied).
In these papers, however, the damping terms in the dielectric tensor
were largely neglected (but see \S 2.5 of HL03). Including
these damping terms can (for certain parameter regimes) modify 
the mode property near the vacuum resonance and give rise
to the phenomenon of mode collapse.

Following Ginzburg (1970), we consider a cold, magnetized plasma
composed of electrons and ions (with charge, mass and number density
given by $-e,~m_e,~n_e$ and $Ze,~M=Am_p,~n_i=n_e/Z$, respectively;
here $Z$ is the charge number and $A$ is the mass number of the ion).
The electrons and ions are coupled by collisions (with the collision frequency
$\nu_{ei}$). We generalize Ginzburg's result by including
the radiative dampings of electrons and ions, with the
damping frequencies $\nu_{re}$ and $\nu_{ri}$, respectively. 
In the coordinate system $XYZ$ with $\bB$ along $Z$, the plasma contribution
to the dielectric tensor is given by 
\footnote{Unlike the expression given in Ginzburg (1970), equations 
(\ref{eq:epsilong}) and (\ref{eq:eta}) are accurate as long as
$\gamma_{ei},~\gamma_{re},~\gamma_{ri}\ll 1$; these accurate
expressions are necessary in order to recover the correct scattering 
cross-sections. Also note that when the damping terms are neglected, 
eqs.~(\ref{eq:epsilong})-(\ref{eq:eta}) agree with eqs.~(2.2)-(2.4) 
of HL03. When the damping terms are included,
the latter equations are incorrect. The reason is that in the 
presence of electron-ion collisions, the electron and ion cannot
be treated as independent particles. The difference between these
two sets of equations is appreciable only for $\omega\lo \omega_{Bi}$.
See Potekhin \& Chabrier (2002) for a discussion.}
\be
\lb\beps^{(p)}\rb_{{\hat Z}={\hat B}}= \lb \begin{array}{ccc}
\varepsilon & ig & 0 \\
-ig & \varepsilon & 0 \\
0 & 0 & \eta
\end{array} \rb, \label{eq:epsij0}
\ee 
where 
\ba
&&\hskip -0.8cm \varepsilon\pm g\simeq 1-{v_e(1+i\gamma_{ri})+v_i(1+
i\gamma_{re})
\over (1+i\gamma_{re}
\pm u_e^{1/2})
(1+i\gamma_{ri}\mp u_i^{1/2})+i\gamma_{ei}^\pm},\label{eq:epsilong}\\
&&\hskip -0.8cm \eta\simeq 1-{v_e\over 1+i(\gamma_{ei}+\gamma_{re})}
-{v_i\over 1+i(\gamma_{ei}+\gamma_{ri})}.
\label{eq:eta}
\ea
In eqs.~(\ref{eq:epsilong}) and (\ref{eq:eta}), we have
defined the dimensionless quantities
\be
u_e={\omega_{Be}^2\over\omega^2}, \quad
u_i={\omega_{Bi}^2\over\omega^2}, \quad
v_e={\omega_{pe}^2\over\omega^2}, \quad
v_i={\omega_{pi}^2\over\omega^2},
\ee
where $\omega_{Be}=eB/(m_ec)$ is the electron cyclotron frequency,
$\omega_{Bi}=ZeB/(Mc)=(Zm_e/Am_p)\omega_{Be}$ is the ion 
cyclotron frequency, $\omega_{pe}=(4\pi n_e e^2/m_e)^{1/2}$
is the electron plasma frequency, and $\omega_{pi}=
(4\pi n_i Z^2e^2/M)^{1/2}=(Zm_e/Am_p)^{1/2}\omega_{pe}$ is the 
ion plasma frequency. 
The dimensionless damping rates
$\gamma_{ei}=\nu_{ei}/\omega$, $\gamma_{re}=\nu_{re}/\omega$, and
$\gamma_{ri}=\nu_{ri}/\omega$ are given by
\ba
\gamma_{ei}&=&
{Z^2n_i e^4\over\hbar\omega^2}\!\left({2\pi\over m_ekT}\right)^{1/2}
\!\!\!(1-e^{-\hbar\omega/kT})\,g_\alpha^{\rm ff}\nonumber\\
&=& 9.2\times 10^{-5}{Z^2\over A}{\rho_1\over T_6^{1/2}E_1^2}(1-e^{-E/kT})
\,g_\alpha^{\rm ff},\label{eqgammaei}\\
\gamma_{re}&=&{2e^2\omega\over 3m_ec^3}=9.5\times 10^{-6}E_1,
\label{eqgammare}\\
\gamma_{ri}&=&{Z^2m_e\over Am_p}\gamma_{re}=5.2\times 10^{-9}{Z^2\over A}E_1,
\ea
where $g_\alpha^{\rm ff}$ is the Gaunt factor. In eq.~(\ref{eq:epsilong}),
\be
\gamma_{ei}^\pm=\gamma_{ei}\left[1\mp u_i^{1/2}(1-Z^{-1})+m_e/(Am_p)\right].
\ee
For Hydrogen plasmas, $\gamma_{ei}^\pm\simeq\gamma_{ei}$.

Including vacuum polarization, the dielectric tensor
$\beps$ and inverse permeability tensor $\bmu^{-1}\equiv \bar\bmu$ 
can be written as $\beps=\beps^{(p)}+\Delta\beps^{(v)}$,
$\bar\bmu=\bI+\Delta\bar\bmu^{(v)}$ (where $\bI$ is the unit tensor).
The vacuum corrections are
\ba
&&\Delta\beps^{(v)}=(a-1)\bI +q\hat\bB\hat\bB,\\
&&\Delta\bar\bmu^{(v)}=(a-1)\bI+m\hat\bB\hat\bB,
\ea
where $a$, $q$, and $m$ are functions of $B$ (the explicit 
expressions are given in LH02 and HL03).
Including the contribution from vacuum polarization, the
dielectric tensor can still be written in the form
of eq.~(\ref{eq:epsij0}), except 
\be
\varepsilon\rightarrow \varepsilon'=\varepsilon+a-1,\qquad
\eta\rightarrow\eta'=\eta+a+q-1.
\ee

With the dielectric tensor and permeability tensor known,
the photon modes can be determined in a straightforward manner. 
In the $xyz$ coordinates with ${\bf k}$ along the $z$-axis and 
lies in the X-Z plane 
(such that $\hat\bB\times {\hatk}=\sin\theta_B\,\hat y$,
where $\theta_B$ is the angle between ${\bf k}$ and $\bB$), we write
the electric field of the mode as $\bE_\pm \propto (iK_\pm,1,iK_{z\pm})$, 
where the ellipticity $K_\pm=-iE_x/E_y$ is given by
\be
K_{\pm}=\beta\pm\sqrt{\beta^2+r},
\label{eqkpm}
\ee
with $r=1+(m/a)\sin^2\theta_B\simeq 1$, and  
the (complex) polarization parameter $\beta$ is 
\be
\beta = -\frac{\varepsilon'^2-g^2-\varepsilon'\eta'
\lp 1+\vpm/\vpa\rp}{2g\eta'}\frac{\sin^2\thetab}{\cos\thetab}.
\label{eq:polarb2}
\ee
The  refractive index $n_\pm$ is given by
\be
n_\pm^2 = \frac{g\eta'}{\vpa\epsilon_{33}}\lp\frac{\varepsilon'}{g}
 + \frac{1}{K_\pm}\cos\thetab\rp, \label{eq:nrefract2}
\ee
where $\epsilon_{33}=\varepsilon'\sin^2\thetab+\eta'\cos^2\thetab$.

The polarization parameter $\polarb$ directly determines the
characteristics of photon normal modes in the medium. For $v_e\ll 1$,
the real part of $\beta$ can be written as ${\rm Re}(\beta)=\beta_0\beta_V$,
where 
\be
\beta_0\simeq \frac{\uel^{1/2}\sin^2\theta_B}{2\cos\theta_B}
\lp 1-\uion\rp, \label{eq:polarb0}
\ee
and 
\be
\beta_V\simeq 1+\frac{\lp\vpq+\vpm\rp\lp 1-\uel\rp}{\uel\vel}.
 \label{eq:polarbvp0}
\ee
The condition ${\rm Re}(\beta)=0$ selects out three critical 
photon energies $E_{Be}=\hbar\omega_{Be},~E_{Bi}=\hbar\omega_{Bi}$ 
and $E_V$ (for a given density),
where the latter is the vacuum resonance energy
\be
E_V\simeq {\hbar\omega_{pe}\over\sqrt{q+m}}.
\ee
We shall concentrate on this vacuum resonance in the remainder of this paper.
Since $E_V$ depends on density, a photon with a given energy $E$
traveling in an inhomogeneous medium encounters the vacuum
resonance at the density $\rho_V$, given by eq.~(\ref{eq:densvp}).
Figure 3 show some examples of the mode properties near the vacuum resonance.

Including the damping terms in the dielectric tensor gives rise to
the phenomenon of mode collapse (see Soffel \etal~1983 and
references therein). This occurs when the two polarization modes become
identical, i.e., $K_+=K_-$ or $n_+=n_-$, which requires $\beta=\pm
i\sqrt{r}\approx\pm i$. Using eq.~(\ref{eq:polarb2}),
we find, for $u_e\gg 1$, $v_e\ll 1$, and neglecting ions
($u_i\ll 1$), 
\be
\polarb \approx \frac{\uel^{1/2}}{2}\frac{\sin^2\thetab}{\cos\thetab}
 \lp\polarbvp-i\gamdampe\rp \label{eq:polarbmcp}
\ee
where $\polarbvp\approx 1-(\vpq+\vpm)/\vel=1-(E/\Evp)^2$
and $\gamma_e=\gamma_{ei}+\gamma_{re}$.
So the condition $\beta=\pm i$ selects out a critical angle
$\theta_B=\theta_{\rm coll}$ at which mode collapse
occurs: $\sin^2\theta_{\rm coll}/\cos\theta_{\rm coll}=
2/(u_e^{1/2}\gamma_e)$.  When the ion effect is included
(i.e., when $u_i\go 1$), no simple expression
for the complex $\beta$ can be obtained, and $\theta_{\rm coll}$
must be calculated numerically (see Fig.~4). Figure 4 shows that 
for ``ordinary'' field strengths ($B_{14}\lo 0.1$), $\theta_{\rm coll}$
is very close to $90^\circ$ for most photon energies (in agreement
with Soffel et al.~1983); for superstrong field strengths 
($B_{14}\go 1$), however, $\theta_{\rm coll}$ can be much smaller
than $90^\circ$. Figure 4 also shows that at $E=E_{Bi}$
(the ion cyclotron energy), the mode collapse point is 
always very close to $\theta_B=90^\circ$.

More generally, the modal description of radiative transfer is
valid only in the limit of large Faraday depolarization, when
the phase shift between the two modes over a mean free path is
significant, i.e., when the condition $|\mbox{Re}(n_+-n_-)|
\gg|\mbox{Im}(n_++n_-)|$ is satisfied (Gnedin \& Pavlov~1974). 
Consider first the situation where the ion effect is neglected 
(i.e., $u_i\ll 1$). For $\uel\gg 1$, we find
\ba
\hskip -0.6cm 
\mbox{Re}(n_+^2-n_-^2) &\approx& -{2\vel\over \uel^{1/2}}\cos\thetab
 \mbox{Re}\lp\sqrt{1+\polarb^2}\rp, \label{eq:mcpreal}\\
\hskip -0.6cm \mbox{Im}(n_+^2+n_-^2) &\approx& \gamdampe\vel\lb\sin^2\thetab
 + {1\over \uel}\lp 1+\cos^2\thetab\rp\rb, \label{eq:mcpimag}
\ea
where $\beta$ is given by eq.~(\ref{eq:polarbmcp}).
For $|\polarbvp|\ga {\rm a~few}\times \gamdampe$
(i.e., for $E$ slightly away from $\Evp$),
we have $|{\rm Re}(n_+^2-n_-^2)|\sim 2v_e u_e^{-1/2}
{\rm max}(\cos\theta_B,u_e^{1/2}\!\beta_V\sin^2\theta_B)$,
thus the condition for Faraday depolarization is satisfied for all $\thetab$.
On the other hand, for $|\polarbvp|\ll\gamdampe$
(i.e., for $E$ extremely close to $\Evp$), the condition
$|\mbox{Re}(n_+-n_-)|\ga|\mbox{Im}(n_++n_-)|$ requires
$(\gamdampe\uel^{1/2}/2)|\sin^2\thetab/\cos\thetab|=|\mbox{Im}(\polarb)|\la 1$;
this condition translates to $\thetab\la\thetamcp$ or
$\thetab\ga\pi-\thetamcp$.
In other words, the breakdown of Faraday depolarization at
the vacuum resonance ($|\polarbvp|\ll\gamdampe$) is restricted
to an angular region of $\thetab$ around $\pi/2$.
For a given $B$ and $\theta_B$, 
to satisfy Faraday depolarization requires
$E\gg\Emcp(B,\theta_B)$, where $\Emcp(B,\theta_B)$
is obtained by setting ${\rm Im}(\beta)=1$.
This result can be generalized to the case where
the ion effect is not negligible (i.e., $u_i\go 1$),
although the simple expressions (\ref{eq:mcpreal})
-(\ref{eq:mcpimag}) are no longer valid.
In general, while mode collapse at the vacuum resonance($\beta_V=0$)
occurs only at two angles $\theta_B=\theta_{\rm coll}$ and
$\pi-\theta_{\rm coll}$, the breakdown of Faraday
depolarization occurs for $\theta_{\rm coll}\lo\theta_B
\lo\pi-\theta_{\rm coll}$. To put it another way,
for a given $B$ and $\theta_B$, there is a range of
energies for which Faraday depolarization breaks down; 
we see from Fig.~5 that this range becomes larger as $B$ increases.

The above analysis shows that in the superstrong field
regime ($B_{14}\go 1$), mode collapse and breakdown
of Faraday depolarization at vacuum resonance
can occur for a wide range of photon energies and
propagation directions. Therefore, a rigorous treatment of radiative 
transfer near the vacuum resonance, in general, requires solving the 
transport equations for the four photon intensity matrix,
or Stokes parameters.  

\section{Equations for the Evolution of Electromagnetic Waves 
and Mode Amplitudes}

An electromagnetic (EM) wave with a given frequency $\omega$ satisfies 
the wave equation
\be
\nabla\times(\bar\bmu\cdot\nabla\times\bE)={\omega^2\over c^2}\,\beps\cdot
\bE.
\label{waveeq}\ee
We assume that the medium is weakly anisotropic so that deflection of
the photon trajectory can be neglected. This requires that the deviation 
of $\beps$ or $\bar\bmu$ from the unit tensor is small. 
Consider a wave propagating in the $z$-direction. 
The external magnetic field is assumed to be constant, and the
medium density varies in space with characteristic length scale
$H_\rho\gg c/\omega$. Let $\bE=e^{ik_0z}\bA$, where $k_0\equiv
\omega/c$. Assuming that $|d\bA/dz|\ll k_0|\bA|$ (geometric
optics approximation), we have
\ba
&& {dA_x\over dz} ={ik_0\over 2a}\bigl[(\epsilon_{11}-a)A_x+\epsilon_{12}A_y+
\epsilon_{13}A_z \bigr],\\
&& {dA_y\over dz} ={ik_0\over 2(a+m\sin^2\theta_B)}
\bigl[\epsilon_{21}A_x \nonumber\\
&&\quad\qquad +(\epsilon_{22}-a-m\sin^2\theta_B)A_y
+\epsilon_{23}A_z \bigr],\\
&&~~A_z= {\epsilon_{13}A_x+\epsilon_{23}A_y\over \epsilon_{33}},
\ea
where $\epsilon_{ij}$ is the component of the dielectric tensor in
the $xyz$ coordinate system (with the $z$-axis along ${\bf k}$).
We shall find that near the vacuum resonance, $\bA$ can vary
on the length scale $1/(k_0|n_+-n_-|)$ (see \S 5.2),
much shorter than $H_\rho$, but the condition
$|d\bA/dz|\ll k_0|\bA|$ is still easily satisfied as long as
$|n_+-n_-|\ll 1$. Using the expressions for $\epsilon_{ij}$, we find that 
$|\epsilon_{13}/\epsilon_{33}|\simeq |(v_e-q)\sin\theta_B\cos\theta_B|$
and $|\epsilon_{23}/\epsilon_{33}|\simeq |v_e u_e^{-1/2}\sin\theta_B/(1-u_i)|$
for $|u_i-1|\go m_e/m_p$ (and $u_e\gg 1$). 
Thus, as long as $v_e\ll 1$ (valid for sufficiently low densities)
and $q\ll 1$ (valid for $B_{14}\ll 600$), the
wave is transverse for most photon energies (an exception
occurs when $E$ is very close to $E_{Bi}$).
We shall adopt this transverse approximation in the following
\footnote{This approximation is also consistent with the 
requirement that the medium be weakly anisotropic.}. 
Since $a-1,~m\ll 1$, the evolution equation for $\bA$ 
simplifies to 
\be
{d\over dz}\left(\begin{array}{c}A_x\\A_y\end{array}\right)
={ik_0\over 2}\left[\begin{array}{cc}
\sigma_{11} & \sigma_{12} \\
\sigma_{21} & \sigma_{22}
\end{array} \right]
\left(\begin{array}{c}A_x\\A_y\end{array}\right)
\label{eqap}\ee
where 
\ba
\sigma_{11}&=&\epsilon_{11}-a
=\varepsilon'\cos^2\theta_B+\eta'\sin^2\theta_B-a,\label{sig11}\\
\sigma_{12}&=&-\sigma_{21}=\epsilon_{12}=ig\cos\theta_B,\label{sig12}\\
\sigma_{22}&=&\epsilon_{22}-a-m\sin^2\theta_B
=\varepsilon'-a-m\sin^2\theta_B.\label{sig22}
\ea
Equation (\ref{eqap}) can be used to study the evolution of 
EM wave amplitude across the vacuum resonance under the most general 
conditions. Equivalently, we can use the photon density matrix 
to study the wave propagation (\S5).  

\subsection{Evolution of Mode Amplitude}

It is instructive to show that when dissipation is neglected, 
eqs.~(\ref{eqap}) can be reduced to the evolution
equations for the mode amplitudes given in LH02.
Let 
\be
\left(\begin{array}{c} A_x\\ A_y\end{array}\right)
=\exp \left[{ik_0\over 4}\int^z(\sigma_{11}+\sigma_{22})dz\right]
\left(\begin{array}{c}\cA_x\\ \cA_y\end{array}\right).
\ee
Equations (\ref{eqap}) reduce to 
\be
{d\over dz}\left(\begin{array}{c}\cA_x\\ \cA_y\end{array}\right)
={ik_0 \Delta n\over 2}\left[\begin{array}{cc}
\cos 2\theta_m & i\sin 2\theta_m \\
-i\sin 2\theta_m & -\cos 2\theta_m
\end{array} \right]
\left(\begin{array}{c}\cA_x\\ \cA_y\end{array}\right)
\label{eqap1}\ee
where $\Delta n$ is the difference of the indices of refraction
for the two modes and $\theta_m$ is the ``mixing angle'', as given by
\be
\Delta n={-i \sigma_{12}\over \sin 2\theta_m},
\quad \tan 2\theta_m =  {-2i \sigma_{12}\over \sigma_{11}-\sigma_{22}}.
\label{deltan}\ee
When dissipation is neglected, $\Delta n$ and $\theta_m$ are real, and
for $u_e\gg 1$, we find
\ba
\Delta n &=&-{v_e\cos\theta_B\over u_e^{1/2}(1-u_i)\sin 2\theta_m},\\
(\tan 2\theta_m)^{-1} &=&
{u_e^{1/2}(1-u_i)\sin^2\theta_B\over 2\cos\theta_B}
\left(1-{q+m\over v_e}\right)\nonumber\\
&=&\beta.
\label{eqbeta3}\ea
The normal mode eigenvectors are $(\cA_x,\cA_y)=(i\cos\theta_m,\sin\theta_m)$
and $(-i\sin\theta_m,\cos\theta_m)$. We define the mode amplitude
$\cA_+$ and $\cA_-$ via
\be
\left(\begin{array}{c} \cA_x\\ \cA_y\end{array}\right)
=\cA_+ \left(\begin{array}{c} i\cos\theta_m\\ \sin\theta_m\end{array}\right)
+\cA_- \left(\begin{array}{c} -i\sin\theta_m\\ \cos\theta_m\end{array}\right).
\label{eqaxay}\ee
The equations for the evolution of $\cA_+$ and $\cA_-$ are then given by
\be
i{d\over dz} \left(\begin{array}{c}\cA_+\\ \cA_-\end{array}\right)
= \left[\begin{array}{cc} -k_0 \Delta n/2 & i\,d\,\theta_m/dz\\
-i\,d\,\theta_m/dz & k_0 \Delta n/2\end{array} \right]
\left(\begin{array}{c}\cA_+\\ \cA_-\end{array}\right),
\label{eqap2}
\ee
which agree with eq.~(15) of LH02. Equations (\ref{eqap1}) and (\ref{eqap2})
have the standard forms of the level crossing problem in
quantum mechanics (Landau 1932; Zener 1932). When 
\be
k_0\left|\Delta n\right|/2\gg \left|{d\,\theta_m/dz}\right|, 
\label{eqadcon}\ee
the polarization vector will evolve
adiabatically (i.e., evolve along the continuous $K_+$ or
$K_-$ curve). Evaluating this condition at vacuum resonance
yields $E\gg E_{\rm ad}$ [see eq.~(\ref{condition}) and LH02 for details]. 
In general, for a linear density profile, 
$\rho=\rho_V(1-z/H_\rho)\simeq \rho_V\exp(-z/H_\rho)$ (for $|z|\ll
H_\rho$), the nonadiabatic jump probability is
given by
\be
P_{\rm jump}=\exp\left[-{(\pi/2)}\left(E/E_{\rm ad}\right)^3\right].
\label{jump}\ee
Thus, in practice, when $E\go 1.3\,E_{\rm ad}$,
the evolution will be highly adiabatic (with $P_{\rm jump}\lo 0.03$).
Figure 6 shows the adiabatic regime $E>E_{\rm ad}$ 
in the $B$--$E$ plane for typical values of $\theta_B$ and
$H_\rho$. [Another way to understand the adiabatic
condition is discussed in \S 5.1; see eq.~(\ref{eqad2})].

\section{Evolution of Stokes Parameters Across Vacuum Resonance}

\subsection{Evolution Equations for Photon Intensity Matrix}

We now study the transfer equation for the photon intensity 
matrix\footnote{It is beyond the scope of this paper to properly review
previous works on transfer equations of polarized radiation in a magnetized
medium; see, e.g., Unno 1956; Dolginov et al.~1970; Dolginov \& Pavlov 1974; 
Zheleznyakov et al.~1974. These papers study the transfer equation
under a variety of different approximations, in different contexts and
with different level of generalities.},
taking into account the birefringence of the plasma and the 
attenuation due to absorption and scattering; we relegate the derivation 
of the source functions to \S6.

The intensity matrix $I_{\alpha\beta}$ of the radiation
is defined as 
\be
I_{\alpha\beta}\equiv C\langle A_\alpha A_\beta^\ast\rangle,
\label{eqiab}\ee
where $\langle\cdots\rangle$ denotes time average
($\alpha,\beta=1,2$ or $x,y$). We choose
the proportional constant $C$ such that $I_{\alpha\beta}$
has the units of specific intensity. The Stokes parameters
$I,Q,U,V$ are related to $I_{\alpha\beta}$ via
\be
\left[I_{\alpha\beta}\right]=
{1\over 2}\left[\begin{array}{cc}
I+Q & U+iV \\
U-iV & I-Q
\end{array} \right].
\ee
Using eq.~(\ref{eqap}), we can derive the evolution equation for
the intensity matrix\footnote{This form of the transfer equation
was derived in Dolginov et al.~(1970) for a general magnetoactive
plasma, although no simple expression for $T_{\alpha\beta}$
was given there.}:
\be
{d\over dz}I_{\alpha\beta}=-{1\over 2}k_0\sum_\gamma\left(T_{\alpha\gamma}
I_{\gamma\beta}+I_{\alpha\gamma}T_{\gamma\beta}^+\right),
\label{eqt}\ee
where 
\be
T_{\alpha\beta}=-i\sigma_{\alpha\beta}.
\ee
Equivalently, the Stokes parameters $\bS=(S_0,S_1,S_2,S_3)^+=(I,Q,U,V)^+$
evolve according to
\be
{d\over dz}
\left(\begin{array}{c}
I\\
Q\\
U\\
V
\end{array} \right)
=k_0
\left[\begin{array}{cccc}
T_0 & T_1 & T_2 & T_3\\
T_1 & T_0 & -\Omega_3 & \Omega_2 \\
T_2 & \Omega_3 & T_0 & -\Omega_1\\
T_3 & -\Omega_2 & \Omega_1 & T_0
\end{array} \right]
\left(\begin{array}{c}
I\\
Q\\
U\\
V
\end{array} \right)
\label{eqstokes}\ee
where
\ba
&& T_0=-{1\over 2}\left(\sigma_{11i}+\sigma_{22i}\right),\\
&& T_1=-{1\over 2}\left(\sigma_{11i}-\sigma_{22i}\right),\\
&& T_2=-{1\over 2}\left(\sigma_{12i}+\sigma_{21i}\right)=0,\\
&& T_3={1\over 2}\left(\sigma_{12r}-\sigma_{21r}\right)=\sigma_{12r},\\
&& \Omega_1={1\over 2}\left(\sigma_{11r}-\sigma_{22r}\right),\\
&& \Omega_2={1\over 2}\left(\sigma_{12r}+\sigma_{21r}\right)=0,\\
&& \Omega_3={1\over 2}\left(\sigma_{12i}-\sigma_{21i}\right)=\sigma_{12i}.
\ea
Here $\sigma_{\alpha\beta r}={\rm Re}(\sigma_{\alpha\beta})$
and $\sigma_{\alpha\beta i}={\rm Im}(\sigma_{\alpha\beta})$.
Following Kubo \& Nagata (1983), equation (\ref{eqstokes}) 
can also be cast into a vector form. Define vectors in the
Poincar\'e sphere
\ba
&&\bs=(S_1/S_0,S_2/S_0,S_3/S_0),\\
&&\bT=(T_1,T_2,T_3),\\
&&\bOmega=(\Omega_1,\Omega_2,\Omega_3).
\ea
Then
\be
{d\over dz}\bs=k_0\left[\bOmega\times\bs+\bs\times (\bT\times\bs)
\right].
\label{eqds}\ee

In the absence of damping, $\bT=0$, and from eq.~(\ref{deltan})
we have\footnote{Heyl and Shaviv (2000) used an equation similar to
eq.~(\ref{eqds}) (with $\bT=0$) to study the polarization evolution of 
EM waves in a pure magnetized vacuum (for which $\theta_m=0$)
with a nonuniform external magnetic field.}
\be
\bOmega=\Delta n \left(\cos 2\theta_m,0,\sin 2\theta_m\right).
\ee
With this, the adiabatic mode evolution discussed in \S 4
can also be understood as follows. Consider a photon initially in the
plus-mode, with $(E_x,E_y)\propto (i\cos\theta_m,\sin\theta_m)$;
it can be represented on the Poincar\'e sphere 
by the vector $\bs=(\cos 2\theta_m,0,\sin 2\theta_m)$. Thus
$\bs$ is initially parallel to $\bOmega$. As the photon propagates in 
an inhomogeneous medium, $\hat\bOmega=\bOmega/\Delta n$ evolves on the
Poincar\'e sphere. If $\hat\bOmega$ changes slowly, $\bs$ will evolve in the
way that it remains almost parallel to $\hat\bOmega$; this requires 
\be
\left|{d(\bOmega/\Delta n)\over dz}
\right|\ll k_0|\bOmega|,
\label{eqad2}\ee
or $2 |d\theta_m/dz|\ll
k_0|\Delta n|$, in agreement with eq.~(\ref{eqadcon}).

For use in \S 6, we define 
\be
\bcI=(I_{11},I_{22},U,V)^+.
\label{eqbci}\ee
The evolution equation can then be written as
\be
{d\over dz}\bcI=-\bcM\cdot \bcI,
\label{eqdi}
\ee
where
\ba
&&\hskip -0.6cm\bigl[\bcM\bigr]=\nonumber\\
&&\hskip -0.3cm k_0
\left[\begin{array}{cccc}
\sigma_{11i} & 0 & {\sigma_{12i}/ 2} & -{\sigma_{12r}/ 2} \\
0 & \sigma_{22i} & -{\sigma_{12i}/ 2}& -{\sigma_{12r}/ 2} \\
-\sigma_{12i} & \sigma_{12i} & {(\sigma_{11i}\!+\!\sigma_{22i})/2}
 &{(\sigma_{11r}\!-\!\sigma_{22r})/ 2}\\
-\sigma_{12r} & -\sigma_{12r} & ({\sigma_{22r}\!-\!\sigma_{11r})/ 2}
& ({\sigma_{11i}\!+\!\sigma_{22i})/ 2}
\end{array}\right]
\ea

\subsection{Numerical Results}

Figure 7 depicts four examples of the evolution of the Stokes parameters 
when monochromatic radiation (of energy $E=0.1,~0.5,~1,~3$~keV, 
respectively) propagates in an inhomogeneous hydrogen plasma with density
profile $\rho=\rho_V(E)\exp(-z/H_\rho)$, where $H_\rho=5~{\rm cm}$.
The vacuum resonance is located at $\rho=\rho_V(E)$
and $z=0$, and the external magnetic field is $B_{14}=1$ and
$\theta_B=45^\circ$. In all cases, the radiation is assumed
to be in the plus-mode and has intensity $I=1$
(arbitrary units) at $z=-0.3$~cm. For $E=3$~keV, the evolution is highly
adiabatic (see Fig.~6), thus the radiation will remain in the plus-mode, 
with $Q$ changing from $1$ (characteristic of O-mode) to $-1$ (characteristic
of X-mode). For $E=0.1$~keV, the evolution is highly non-adiabatic,
and the four Stokes parameters remain almost 
constant across the vacuum resonance. For $E=0.5$~keV and $1$~keV,
we have the intermediate situation where partial mode coupling/conversion 
takes place at the resonance.

The oscillation of $Q,~U,~V$ after the radiation crosses
the vacuum resonance can be understood as follows. 
Since the modes are orthorgonal and form a complete set away from the
resonance, we can write the radiation field after the resonance as 
\be
\bE=\left(\!\!\begin{array}{c} A_x\\ A_y\end{array}\!\!\right)e^{ik_0z}=
A_+ \left(\!\!\begin{array}{c} i\cos\theta_m\\ \sin\theta_m\end{array}
\!\!\right)
+A_- \left(\!\!\begin{array}{c} -i\sin\theta_m
\\ \cos\theta_m\end{array}\!\!\right),
\label{eqax1}\ee
where $\theta_m$ is real (i.e., the damping terms are neglected in
calculating the modes). The corresponding Stokes parameters are 
\ba
&& I=|A_+|^2+|A_-|^2,\\
&& Q=\cos 2\theta_m\left(|A_+|^2-|A_-|^2\right) \nonumber\\
&&\qquad -2\sin 2\theta_m |A_+| |A_-|\cos\Delta\phi,\\
&& U=-2 |A_+| |A_-|\sin \Delta\phi,\\
&& V=\sin 2\theta_m \left(|A_+|^2-|A_-|^2\right)\nonumber\\
&&\qquad +2\cos 2\theta_m |A_+| |A_-| \cos\Delta\phi,
\ea
where $|A_\pm|\propto \exp(-\rho\kappa_\pm z/2)$ is the mode amplitude
($\kappa_\pm$ is the opacity) and $\Delta\phi=
\Delta\phi_0+k_0\int^z (n_+-n_-)\,dz$ is the phase difference
between the modes ($\Delta\phi_0$ is the initial phase shift).
Clearly, $Q,~U,~V$ oscillate on the length scale $\sim 1/[k_0|n_+-n_-|]$.
Also note that away from the resonance, $\theta_m\rightarrow 0$ or
$\pi/2$, so that the oscillation of $Q$ diminishes rapidly after the 
resonance.

The results of Fig.~7 are obtained using eq.~(\ref{eqdi}). We can 
obtain the same results by integrating eq.~(\ref{eqap}) and calculating
the Stokes parameters from eq.~(\ref{eqiab}); this requires
many more integration steps than using eq.~(\ref{eqdi}) directly to achieve
the same numerical accuracy. Since the photon modes are well-defined away
from the vacuum resonance, we can calculate the mode amplitudes
$A_+$ and $A_-$ from $A_x$ and $A_y$ [from integration of eq.~(\ref{eqap})]
using eq.~(\ref{eqax1}). Figure 8 shows the evolution of the
ratio $|A_+|/|A|$, where $|A|=(|A_+|^2+|A_-|^2)^{1/2}$. Using 
eq.~(\ref{jump}), we find that $P_{\rm jump}=1-3.2\times 
10^{-7},~0.84,~0.27,~6.5\times 10^{-7}$ for $E=0.1,~0.5,~1,~3$~keV, 
respectively.
Starting with $|A_+|/|A|=1$ before resonance, this implies
$|A_+|/|A|=(1-P_{\rm jump})^{1/2}=5.7\times 10^{-4},~0.40,~0.85,~1$
after the resonance, in agreement with the numerical results shown 
in Fig.~8. Note that except for the $E=0.1$~keV case, the three other cases
all lie in the large Faraday depolarization regime (see Figs.~4,~5),
and the agreement between Fig.~8 and eq.~(\ref{jump}) is not surprising.

Figure 9 shows another example of the evolution of polarized radiation,
with $E=1.7$~keV, $B_{14}=5$, $\theta_B=45^\circ$ and $H_\rho=5$~cm.
For this case, $\beta=1.066\,i$ at $\rho=\rho_V$, thus the two modes
nearly collapse at the vacuum resonance. Nevertheless, we find that 
eq.~(\ref{jump}) provides an accurate description for the evolution
across the resonance: after the resonance, $|A_-|=P_{\rm jump}^{1/2}=0.93$
and decays slowly, while the plus-mode damps quickly because of the 
large opacity.

We note the distance over which resonant evolution occurs is 
much smaller than the the photon mean free path (see Figs.~7-9).
We can estimate the
width of the resonance as follows. From eq.~(\ref{eqbeta3}) we have
\be
\beta=
{u_e^{1/2}(1-u_i)\sin^2\theta_B\over 2\cos\theta_B}\,
\left(1-{\rho_V/\rho}\right).
\ee
At the resonance $\beta=0$. If we define the edge of the resonance region
using the consition $|\beta|=\beta_{ed}\go$~a few, 
we find the half-width of the resonance to be given by (LH02)
\ba
&&(\Delta z)_V={2\beta_{ed} H_\rho\cos\theta_B\over
u_e^{1/2}|1-u_i|\sin^2\theta_B}\nonumber\\
&&~\simeq 1.73\times 10^{-3}\beta_{ed}
|1-u_i|^{-1}{\cos\theta_B\over\sin^2\theta_B}
\left({E_1\over B_{14}}\right)H_\rho.
\ea
On the other hand, the imaginary parts of the indices of refraction of the two
modes are given by 
\ba
&&{\rm Im}(n_+^2)\simeq \gamma_ev_e\sin^2\theta_B\cos^2\theta_m,\\
&&{\rm Im}(n_-^2)\simeq \gamma_ev_e\sin^2\theta_B\sin^2\theta_m,
\ea
[valid for $u_e\gg 1$ and $u_i\ll 1$; cf. eq.~(\ref{eq:mcpimag})]. 
The mean free path $l$ is given by $l^{-1}=(\omega/c){\rm Im}(n^2)$,
and thus
$l\ge c/(\omega \gamma_e v_e)$ (the right-hand side being the mean free path 
at $B=0$). When collisional damping dominates, $\gamma_e\simeq \gamma_{ei}
\gg \gamma_{re}$ [see eqs.~(\ref{eqgammaei})-(\ref{eqgammare})], we have
\be 
l\ge 0.26\,{AT_6^{1/2}E_1^3\over Z^2\rho_1^2(1-e^{-E/kT})g_\alpha^{\rm ff}}
~{\rm cm}.
\ee
When scattering dominates, $l\ge 2.5\rho_1^{-1}$~cm. 
Thus the relation $l\gg (\Delta z)_V$ is almost always satisfied.

\section{Transfer Equations for Stokes Parameters}

Including the source term, the transfer equation reads
[cf. eq.~(\ref{eqdi})]
\be
{d\over ds}\bcI=\hatk\cdot\nabla\bcI
=-\bcM\cdot \bcI+\bcS_{\rm em}+\bcS_{\rm sc},
\ee
where we have used $ds$ (instead of $dz$) to specify the derivative
along the ray. We derive the source functions $\bcS_{\rm em}$ and 
$\bcS_{\rm sc}$ below.

\subsection{Source Function: Emission} 

The emissivity $\bcS_{\rm em}$
can be derived by considering the radiation field
in thermodynamic equilibrium with matter, for which
$\bcI=(B_\nu/2,B_\nu/2,0,0)^+$. Note that under  this situation,
$\bcS_{\rm sc}$ should completely cancel out the scattering 
contribution to $\bcM\cdot\bcI$ (see \S 6.2.2). 
Thus
\be
\bcS_{\rm em}={k_0 B_\nu \over 2}\,\bigl(\sigma_{11i},\sigma_{22i},
0,-2\sigma_{12r}\bigr)^+_{\rm em}.
\label{sem}\ee
This is equivalent to adding to the RHS of eq.~(\ref{eqstokes}) a source term
\be
-k_0 B_\nu\,\bigl(T_0,T_1,T_2,T_3\bigr)^+_{\rm em}.
\label{sem2}\ee
This is also the same as adding to the RHS of eq.~(\ref{eqt}) a term
\be
{1\over 4}k_0 B_\nu \left(T_{\alpha\beta}+T_{\alpha\beta}^+\right)_{\rm em}.
\label{sem3}\ee
The subscript ``em'' in  eqs.~(\ref{sem})-(\ref{sem3})
implies that the terms proportional to $\gamma_{re}=\gamma_{ri}$ should 
be excluded in evaluating these equations, since 
these terms are related to scattering contributions (\S 6.2.2).
Equations (\ref{sig11})-(\ref{sig22}) yield
\ba
&&\sigma_{11i}=\varepsilon_i\cos^2\theta_B+\eta_i\sin^2\theta_B,
\label{sig11i}\\
&&\sigma_{22i}=\varepsilon_i,\label{sig22i}\\
&& \sigma_{12r}=-g_i\cos\theta_B.\label{sig12r}
\ea
From eqs.~(\ref{eq:epsilong}) and (\ref{eq:eta}), we find
\ba
&&\hskip -0.5cm
\varepsilon_i\pm g_i={\rm Im}(\varepsilon\pm g)=\nonumber\\
&&\Lambda_\pm\left[v_e\gamma_{ei}^\pm+(1\mp u_i^{1/2})^2v_e\gamma_{re}
+(1\pm u_e^{1/2})^2v_i\gamma_{ri}\right],\label{epsiloni}\\
&&\hskip -0.5cm
\eta_i={\rm Im}(\eta)=v_e\gamma_{ei}+v_e\gamma_e+v_i\gamma_i,
\label{etai}\ea
where
\be
\Lambda_\pm=\left[(1\pm u_e^{1/2})^2(1\mp u_i^{1/2})^2+\gamma_\pm^2
\label{eqlambda}\right]^{-1}\ee
and
\be
\gamma_\pm=\gamma_{ei}^\pm+(1\pm u_e^{1/2})\gamma_{ri}
+(1\mp u_i^{1/2})\gamma_{re}.
\ee
Thus eq.~(\ref{sem}) becomes
\be
\hskip -0.3cm  \bcS_{\rm em}={k_0v_e B_\nu\over 4}
\!\left[\begin{array}{c}
{\cos^2\!\theta_B}\left(\Lambda_+\gamma_{ei}^+
+\Lambda_-\gamma_{ei}^-\right)+2\gamma_{ei}\sin^2\!\theta_B\\
\Lambda_+\gamma_{ei}^+ +\Lambda_-\gamma_{ei}^-\\
0\\
2\left(\Lambda_+\gamma_{ei}^+ -\Lambda_-\gamma_{ei}^-\right)\cos\theta_B
\end{array} \right].
\label{sem4}
\ee

Note that the terms in eq.~(\ref{sem4}) 
can be easily related to the free-free opacity:
in magnetic fields, a photon of a certain polarization has an absorption
opacity given by 
\be
\kappa^{\rm ff}=\kappa_+|e_+|^2 + \kappa_-|e_-|^2
+\kappa_0|e_0|^2,
\ee
where $e_0=e_Z$, $e_\pm=(e_X\pm i e_Y)/\sqrt{2}$ are the 
spherical components (with $\bB$ along the $Z$-axis)
of the photon's unit polarization vector ${\bf e}$, and 
\be
\kappa_\pm ={k_0\over\rho} v_e\gamma_{ei}^\pm\Lambda_\pm,\quad
\kappa_0 ={k_0\over\rho} v_e\gamma_{ei}.
\label{opacity}\ee
Equation (\ref{opacity}) agrees with (and generalizes to include the $Z>1$ 
case) the result of Potekhin \& Chabrier (2002).

\subsection{Source Function: Scattering}

We first derive the source function due to electron scattering.
The contribution from ion scattering can be easily added (see below).

Consider an incident EM wave with electric field
$\bE'(\br,t)=\bE' e^{i\bk'\cdot\br-i\omega t}$, where
$\bE'=E_x'\hate_\xp+E_y'\hate_\yp$.
In the coordinate system $XYZ$ with $\bB$ along the $Z$-axis,
the direction of the wave vector $\bk'=k_0\hatk'$ is specified by the polar
angles $\theta_B'$ and $\phi_B'$. We define the directions of the
unit vectors $\hate_\xp$ and $\hate_\yp$ such that $\hate_\xp$
lies in the meridian plane and points in the direction of
increasing $\theta_B'$, and $\hate_\xp\times\hate_\yp=\hatk'$.
Similarly, the scattered wave is denoted by 
$\bE (\br,t)=\bE\, e^{i\bk\cdot\br-i\omega t}$,  where
$\bE=E_x\hate_x+E_y\hate_y$, and the polar angles for 
$\hatk$ are $\theta_B$ and $\phi_B$. We also define
unit vectors
\be
\hate_\pm=(\hate_X\pm i\hate_Y)/\sqrt{2},\quad
\hate_0=\hate_Z.
\ee
We can write the incident wave as 
\be
\bE'=\sum_{\alpha=0,\pm}
E_\alpha' \hate_\alpha^\ast,\quad {\rm with}~~
E_\alpha'=\hate_\alpha\cdot \bE'.
\ee
(Here and henceforth, $\sum_\alpha$ refers to sum over
$\alpha=0,\pm$).
The scattered wave is given by the standard dipole formula 
$\bE=\hatk\times (\hatk\times\ddot {\bf d})/(c^2r)$, with
\be
\ddot{\bf d}=c^2r_e\sum_{\alpha}(1+\alpha u_e^{1/2})^{-1}E_\alpha'
\hate_\alpha^\ast,
\ee
where $r_e=e^2/(m_ec^2)$.
Thus the Jones matrix relating $\bE'$ and $\bE$ is given by
\be
\left(\begin{array}{c}
E_x\\
E_y\end{array}\right)
=-{r_e\over r}
\left[\begin{array}{cc}
a & b\\
c & d\end{array}\right]
\left(\begin{array}{c}
E_x'\\
E_y'\end{array}\right),
\ee
where 
\ba
&&a=\sum_\alpha (1+\alpha u_e^{1/2})^{-1}(\hate_x\cdot\hate_\alpha^\ast)
(\hate_\alpha\cdot\hate_\xp),\label{eqna}\\
&&b=\sum_\alpha (1+\alpha u_e^{1/2})^{-1}(\hate_x\cdot\hate_\alpha^\ast)
(\hate_\alpha\cdot\hate_\yp),\label{eqnb}\\
&&c=\sum_\alpha (1+\alpha u_e^{1/2})^{-1}(\hate_y\cdot\hate_\alpha^\ast)
(\hate_\alpha\cdot\hate_\xp),\label{eqnc}\\
&&d=\sum_\alpha (1+\alpha u_e^{1/2})^{-1}(\hate_y\cdot\hate_\alpha^\ast)
(\hate_\alpha\cdot\hate_\yp).\label{eqnd}
\ea
Using $\hate_x=\cos\theta_B(\cos\phi_B\hate_X+\sin\phi_B\hate_Y)
-\sin\theta_B\hate_Z$ and $\hate_y=-\sin\phi_B\hate_X+\cos\phi_B\hate_Y$
(and similarly for $\hate_\xp$ and $\hate_\yp$), we find
\ba
&&a=\cos\theta_B'\cos\theta_B\,\Cdp+\sin\theta_B'\sin\theta_B,\label{eqa}\\
&&b=\cos\theta_B \,\Sdp,\label{eqb}\\
&&c=-\cos\theta_B' \,\Sdp,\label{eqc}\\
&&d=\Cdp,\label{eqd}
\ea
where\footnote{To avoid divergence at $u_e=1$, one can replace
$(1+\alpha u_e^{1/2})$ by $(1+i\gamma_e+\alpha u_e^{1/2})$ (where
$\gamma_e$ is the dimensionless electron damping rate) in all 
relevant equations. Similarly, one can replace $(1+\alpha u_i^{1/2})$
by $(1+i\gamma_i+\alpha u_i^{1/2})$ to avoid divergence at $u_i=1$.
For simplicity, we do not include these damping terms explicitly
in the equations of this subsection.}
\ba
&&\Cdp={1\over 2}\left(
{e^{i\Delta\phi}\over 1-u_e^{1/2}}
+{e^{-i\Delta\phi}\over 1+u_e^{1/2}}\right),\\
&&\Sdp={1\over 2i}\left(
{e^{i\Delta\phi}\over 1-u_e^{1/2}}
-{e^{-i\Delta\phi}\over 1+u_e^{1/2}}\right),
\ea
with $\Delta\phi\equiv\phi_B-\phi_B'$. Using the definition
of $\bcI$ [see eq.~(\ref{eqbci})], we can relate $\bcI(\bk)$ 
for the scattered radiation  to $\bcI'=\bcI(\bk')$ for the incident
radiation. Integrating over all possible incident directions $\hatk'$,
we obtain the source function due to electron scattering:
\be
\bcS_{\rm sc}^{(e)}(\bk)=n_e r_e^2\int\!\! d\Omega'\,\,{\bf R}^{(e)}
(\bk'\rightarrow
\bk)\cdot\bcI(\bk'),
\ee
where $d\Omega'=d^2\hatk'=d\cos\theta_B'd\phi_B'$, and the scattering phase
matrix ${\bf R}^{(e)}$ is given by
\ba
&&\hskip -0.8cm {\bf R}^{(e)}(\bk'\rightarrow\bk)=\nonumber\\
&&\hskip -0.6cm \left[\begin{array}{cccc}
|a|^2 & |b|^2 & (a^\ast b)_r & (a^\ast b)_i\\
|c|^2 & |d|^2 & (c^\ast d)_r & (c^\ast d)_i\\
2(a^\ast c)_r & 2(b^\ast d)_r & (a^\ast d+b^\ast c)_r & (a^\ast d-b^\ast c)_i\\
-2(a^\ast c)_i & -2(b^\ast d)_i & -(a^\ast d+b^\ast c)_i &
(a^\ast d-b^\ast c)_r
\end{array}\right]
\label{eqr}\ea
In eq.~(\ref{eqr}), we have used the notation
$(a^\ast b)_r={\rm Re}(a^\ast b)$ and
$(a^\ast b)_i={\rm Im}(a^\ast b)$, etc.

The source function due to photon-ion scattering can be similarly 
written as
\be
\bcS_{\rm sc}^{(i)}(\bk)=n_i r_i^2\int\!\! d\Omega'\,\,{\bf R}^{(i)}
(\bk'\rightarrow \bk)\cdot\bcI(\bk'),
\ee
where $n_i$ is the ion number density, $r_i=(Ze)^2/(Am_pc^2)$, and
${\bf R}^{(i)}$ has the same form as eq.~(\ref{eqr}), except that 
in the expressions for $a,b,c,d$ [eqs.~(\ref{eqa})-(\ref{eqd})] 
one should replace $\Cdp$ and $\Sdp$ by 
\ba
&&\Cidp={1\over 2}\left(
{e^{i\Delta\phi}\over 1+u_i^{1/2}}
+{e^{-i\Delta\phi}\over 1-u_i^{1/2}}\right),\\
&&\Sidp={1\over 2i}\left(
{e^{i\Delta\phi}\over 1+u_i^{1/2}}
-{e^{-i\Delta\phi}\over 1-u_i^{1/2}}\right).
\ea
It is easy to check the source functions derived above reduce
to the appropriate zero-field results (Chandrasekhar 1960)
when $B\rightarrow 0$.

\subsubsection{Special Case: Axisymmetry}

When $\bcI$ is independent of $\phi_B$ [as is the case
when $\bB$ is perpendicular to the (local) stellar surface],
the source function $\bcS_{\rm sc}^{(e)}$ reduces to 
\be
\bcS_{\rm sc}^{(e)}(\bk)=2\pi n_e r_e^2 \int\!d\mu'\,\,
\bar{\bf R}^{(e)}
(\bk'\rightarrow \bk)\cdot\bcI(\bk'),
\ee
where $\mu'=\cos\theta_B'$. 
The azimuthal average of ${\bf R}^{(e)}$ is given by
\be
\bar{\bf R}^{(e)}(\bk'\rightarrow\bk)=
\left[\begin{array}{cccc}
\langle |a|^2\rangle & \mu^2F_e& 0 & \mu'\mu^2 G\\
\mu'^2F_e  & F_e & 0 & \mu'G_e \\
0 & 0 & 0 & 0\\
2\mu'^2\!\mu G_e & 2\mu G_e & 0 & 2\mu'\!\mu F_e,
\end{array}\right]
\label{eqrbar}\ee
where $\mu=\cos\theta_B$ and
\be
\langle |a|^2\rangle=(\mu'\mu)^2F_e+(1-\mu'^2)(1-\mu^2).
\label{eqa2}\ee
The functions $F_e$ and $G_e$ in eqs.~(\ref{eqrbar})-(\ref{eqa2}) are 
\ba
&&\hskip -0.2cm 
F_e=\bigl\langle |C^{(e)}(\Delta\phi)|^2\bigr\rangle=\bigl\langle 
|S^{(e)}(\Delta\phi)|^2\bigr\rangle={1+u_e\over 2(1-u_e)^2},\\
&&\hskip -0.2cm
G_e=i\Bigl\langle C^{(e)}(\Delta\phi)\left[S^{(e)}(\Delta\phi)
\right]^\ast\Bigr\rangle
=-{u_e^{1/2}\over (1-u_e)^2}.
\ea

The source function $\bcS_{\rm sc}^{(i)}$ due to photon-ion scattering is 
\be
\bcS_{\rm sc}^{(i)}(\bk)=2\pi n_i r_i^2 \int\!d\mu'\,\,
\bar{\bf R}^{(i)}
(\bk'\rightarrow \bk)\cdot\bcI(\bk'),
\ee
where $\bar{\bf R}^{(i)}$ has the same form as eq.~(\ref{eqrbar}), except
that the functions $F_e$ and $G_e$ are replaced by
\be
F_i={1+u_i\over 2(1-u_i)^2},\qquad
G_i={u_i^{1/2}\over (1-u_i)^2}.
\ee

\subsubsection{Special Case: Thermodynamic Equilibrium}

When the radiation field is in complete thermodynamic equilibrium 
with matter, $\bcI=(B_\nu/2,B_\nu/2,0,0)^+$, and the source 
functions due to electron and ion scatterings become
\ba
&&\bcS_{\rm sc}^{(e)}(\bk)=n_e\sigma_T B_\nu
\left[\begin{array}{c}
\mu^2F_e+(1-\mu^2)/2\\
F_e\\
0\\
2\mu G_e
\end{array}\right],\\
&&\bcS_{\rm sc}^{(i)}(\bk)=n_i\sigma_T^{(i)} B_\nu
\left[\begin{array}{c}
\mu^2F_i+(1-\mu^2)/2\\
F_i\\
0\\
2\mu G_i
\end{array}\right],
\ea
where $\sigma_T=8\pi r_e^2/3$ and $\sigma_T^{(i)}=8\pi r_i^2/3$.
On the other hand, the sink term due to scattering is
\be
\left(\bcM\cdot\bcI\right)_{\rm sc}
={k_0 B_\nu \over 2}\,\bigl(\sigma_{11i},\sigma_{22i},
0,-2\sigma_{12r}\bigr)^+_{\rm sc},
\ee
where the subscript ``sc'' implies that one should keep only the terms 
proportional to $\gamma_{re}$ and $\gamma_{ri}$ in
eqs.~(\ref{epsiloni})-(\ref{etai}) (the terms proportional
to $\gamma_{ei}$ are due to absorption). 
Neglecting the damping $\gamma_\pm^2$ in eq.~(\ref{eqlambda}), we find
$\varepsilon_i=2(v_e\gamma_{re}F_e+v_i\gamma_{ri}F_i)$,
$g_i=2(v_e\gamma_{re}G_e+v_i\gamma_{ri}G_i)$, and
$\eta_i=v_e\gamma_{re}+v_i\gamma_{ri}$. Substituting these
into eqs.~(\ref{sig11i})-(\ref{sig12r}) and using
\be
v_e\gamma_{re}=n_e\sigma_T/k_0,\qquad
v_i\gamma_{ri}=n_i\sigma_T^{(i)}/k_0,
\ee
we can easily show that $\bcS_{\rm sc}^{(e)}+\bcS_{\rm sc}^{(i)}
-\left(\bcM\cdot\bcI\right)_{\rm sc}=0$ is satisfied, as it should.

\section{Conclusion}

The main results of this paper can be summarized as follows.

1. We presented a physical discussion of the effect
of vacuum polarization on radiative transfer in strong magnetic fields
and thermal radiation from magnetars (\S 2). 
This substantiated and clarified our previous results (LH02 and HL03):
vacuum polarization depletes the high-energy tail of the 
thermal spectrum and suppresses spectral lines. 

2. We studied the mode properties and examined several issues 
associated with the resonance due to vacuum polarization in 
a magnetized electron-ion plasma (\S 3).  
We showed that for superstrong magnetic fields, mode collapse and the breakdown
of Faraday depolarization occur near the vacuum resonance for a wide range of
propagation directions. Therefore a rigorous treatment of 
radiative transfer in magnetar atmospheres must go beyond the modal description
of the radiation.

3. We studied the propagation of polarized radiation in 
the inhomogeneous atmosphere of magnetars (\S 4 and \S5). 
We found that even with mode collapse and breakdown of
Faraday depolarization, the evolution of polarized X-rays across
the vacuum resonance can be described by a jump probability from one 
mode into another.

4. We derived the general transfer equations for the photon
intensity matrix (Stokes parameters) in the magnetized plasma-vacuum
medium (\S6). Explicit expressions for the source functions due to
emission and scatterings were obtained. These transfer equations
are useful for accurate calculations of radiation/polarization 
spectra from magnetars.

\acknowledgments
We thank G. Pavlov and A. Potekhin for useful 
discussions/suggestions, and P. Goldreich for 
several helpful remarks.
This work is supported in part by NASA 
grants NAG 5-8484 and NAG 5-12034, NSF grant AST 9986740, and
a fellowship from the A.P. Sloan foundation.


\clearpage
\begin{figure*}
\plotone{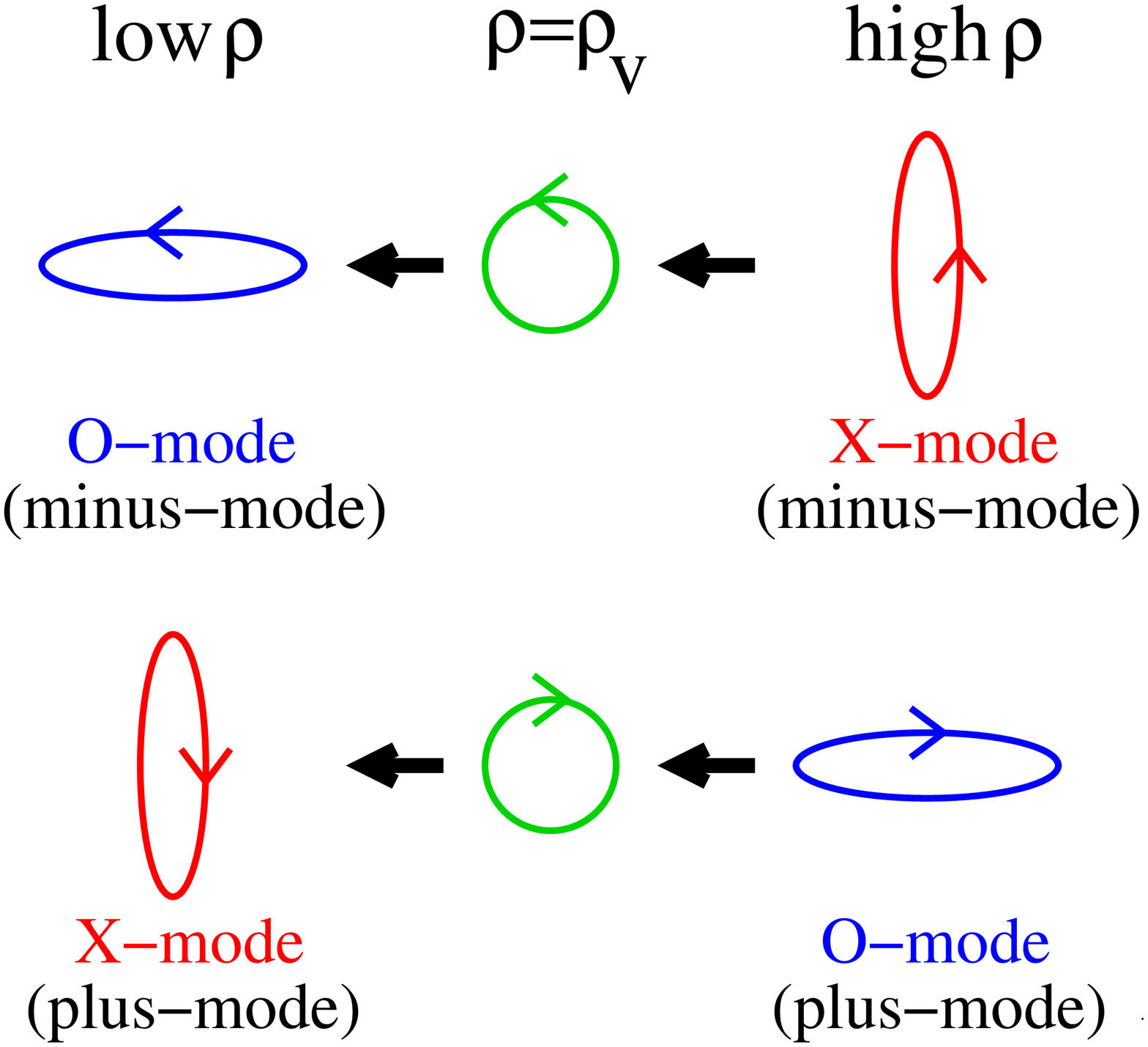}
\vskip 1cm
\caption{A schematic diagram illustrating mode conversion due to
vacuum polarization: As a minus-mode (plus-mode) photon, which manifests 
as X-mode (O-mode) at high density (for $E>E_{Bi}$), traverses the vacuum
resonance density $\rho_V$, it will stay as minus-mode (plus-mode) 
and become O-mode (X-mode) at low density if the adiabatic condition is
satisfied (see also the left panels of Fig.~3).
In this evolution, the polarization ellipse rotates $90^\circ$
and the photon opacity changes significantly.
\label{fig1}}
\end{figure*}

\begin{figure*}
\plotone{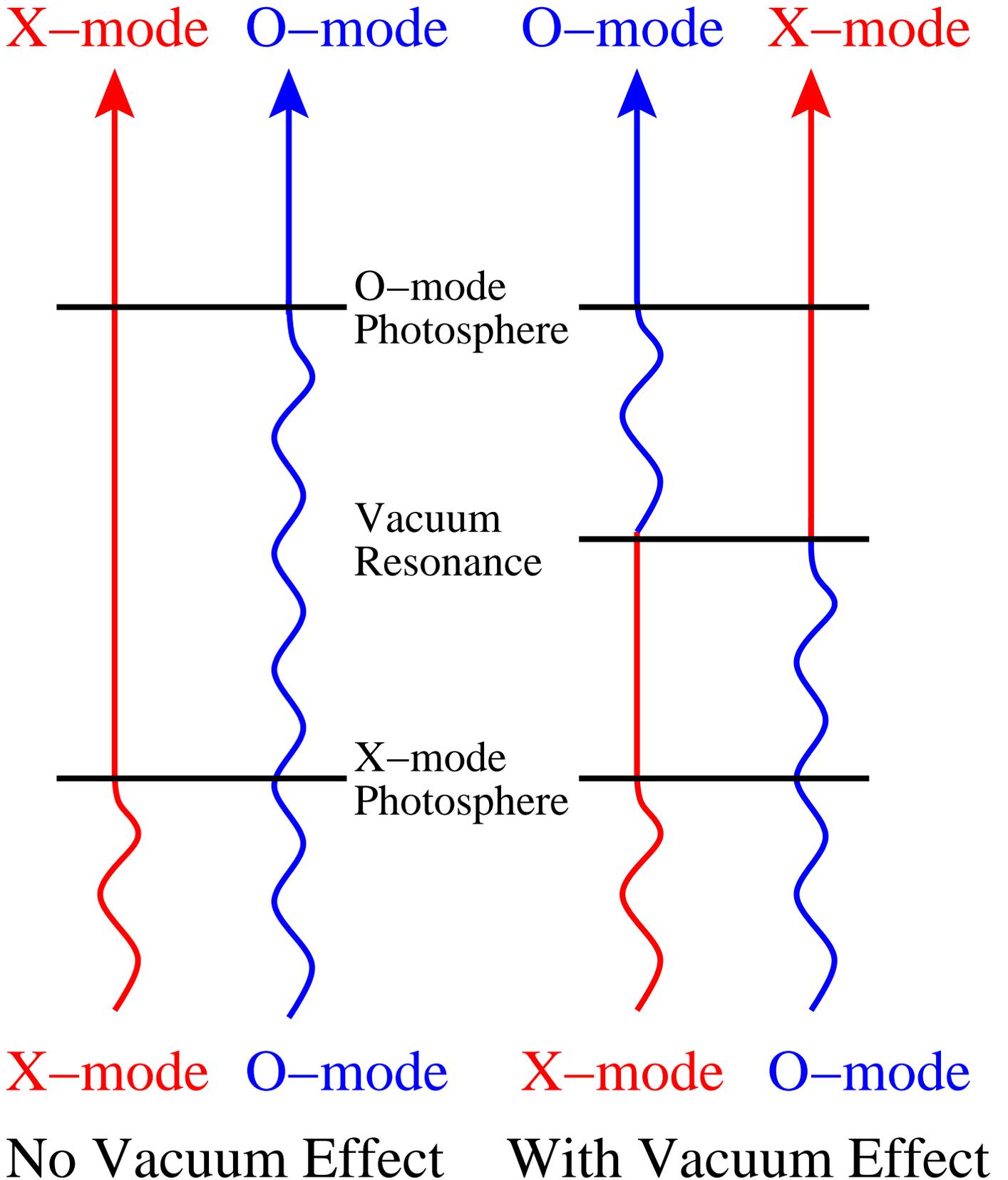}
\vskip 1cm 
\caption{
A schematic diagram illustrating how vacuum polarization-induced 
mode conversion affects radiative transfer in a magnetar atmosphere.
When the vacuum polarization effect is turned off, the X-mode 
photosphere (where optical depth $\sim 1$) lies deeper than the O-mode. 
With the vacuum polarization effect included, the X-mode effectively
decouples (emerges) from the atmosphere at the vacuum resonance,
which lies at a lower density than the (original) X-mode photosphere.
\label{fig2}}
\end{figure*}

\begin{figure*}
\plotone{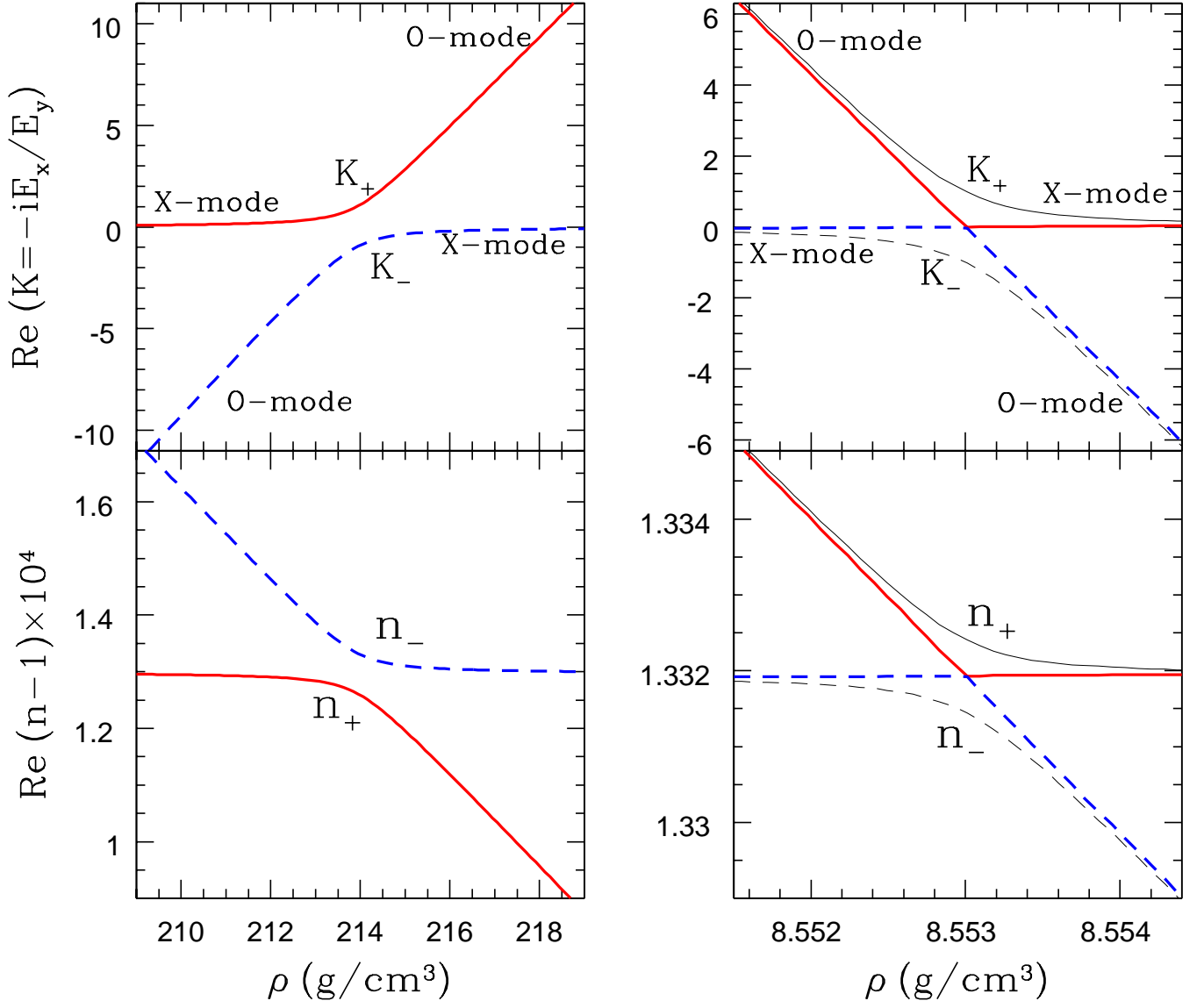}
\caption{The polarization ellipticities $K$ [upper panels; 
see eq.~(\ref{eqkpm})] and refractive indices (lower panels) of the 
photon modes as functions of density near vacuum resonance for $B=5\times
10^{14}$~G, $\theta_B=45^\circ$, and $Y_e=1$. The left panels are for
$E=5$~keV, and the right panels for $E=1$~keV. On the right panels,
the light curves show the results when damping is neglected, while
the heavy lines include damping (we set $T=5\times 10^6$~K in evaluating
$\gamma_{ei}$). On the left panels, the results are 
indistinguishable with and without damping. Note that the correspondence
between the plus-, minus-modes and the X-, O-modes depends on the density
and the photon energy ($E>E_{Bi}$ vs. $E<E_{Bi}$).
\label{fig3}}
\end{figure*}

\begin{figure*}
\plotone{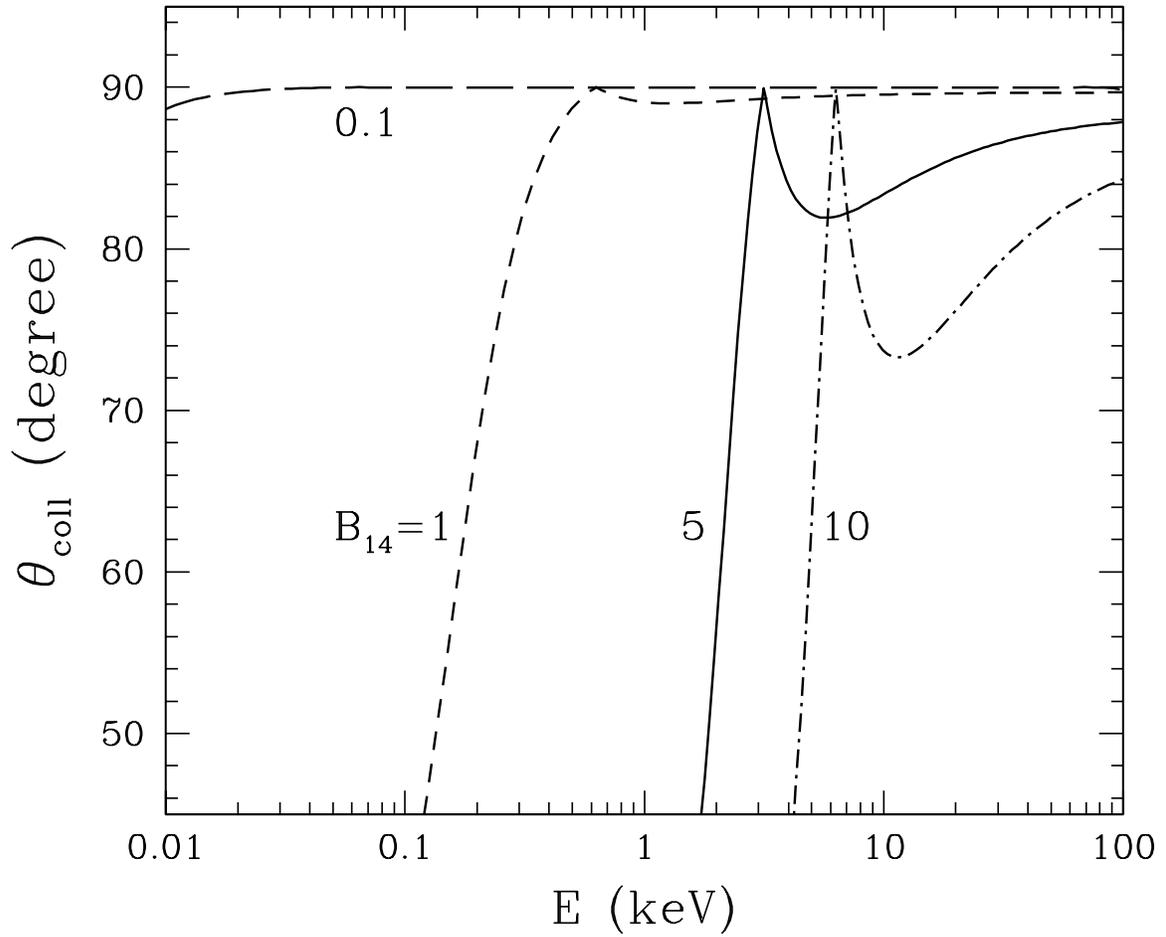}
\caption{The critical angle $\theta_B=\theta_{\rm coll}$
for mode collapse at vacuum resonance 
as a function of photon energy. The four curves correspond
to $B_{14}=0.1,~1,5,~10$. Note that for $E=E_{Bi}$ (the ion cyclotron
energy), $\theta_{\rm coll}$ is very close to $90^\circ$. 
For $\theta_{\rm coll}< \theta_B< 180^\circ-\theta_{\rm coll}$,
the modes are highly nonorthogonal and the condition 
for Faraday depolarization breaks down (see also Fig.~5).
\label{fig4}}
\end{figure*}

\begin{figure*}
\plotone{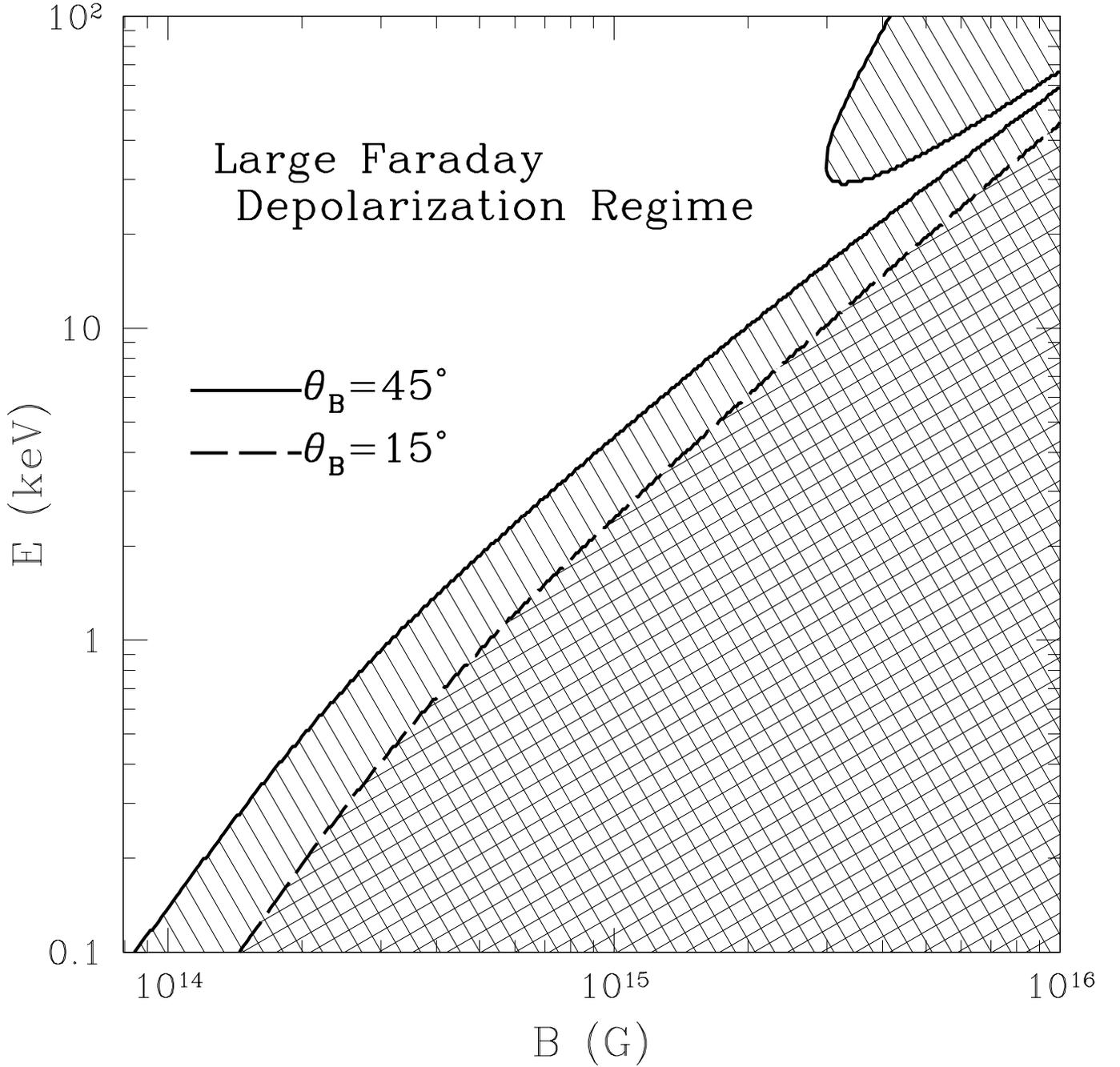}
\caption{The critical photon energy $E=E_{\rm Far}(B,\theta_B)$
for the breakdown of Faraday depolarization. For $E\gg E_{\rm Far}$
(the unshaded region), the condition for large Faraday depolarization is
satisfied. Note that the boundary $E=E_{\rm far}(B,\theta_B)$
also corresponds to $\theta_B=\theta_{\rm coll}(E,B)$ (see Fig.~4).
\label{fig5}}
\end{figure*}

\begin{figure*}
\plotone{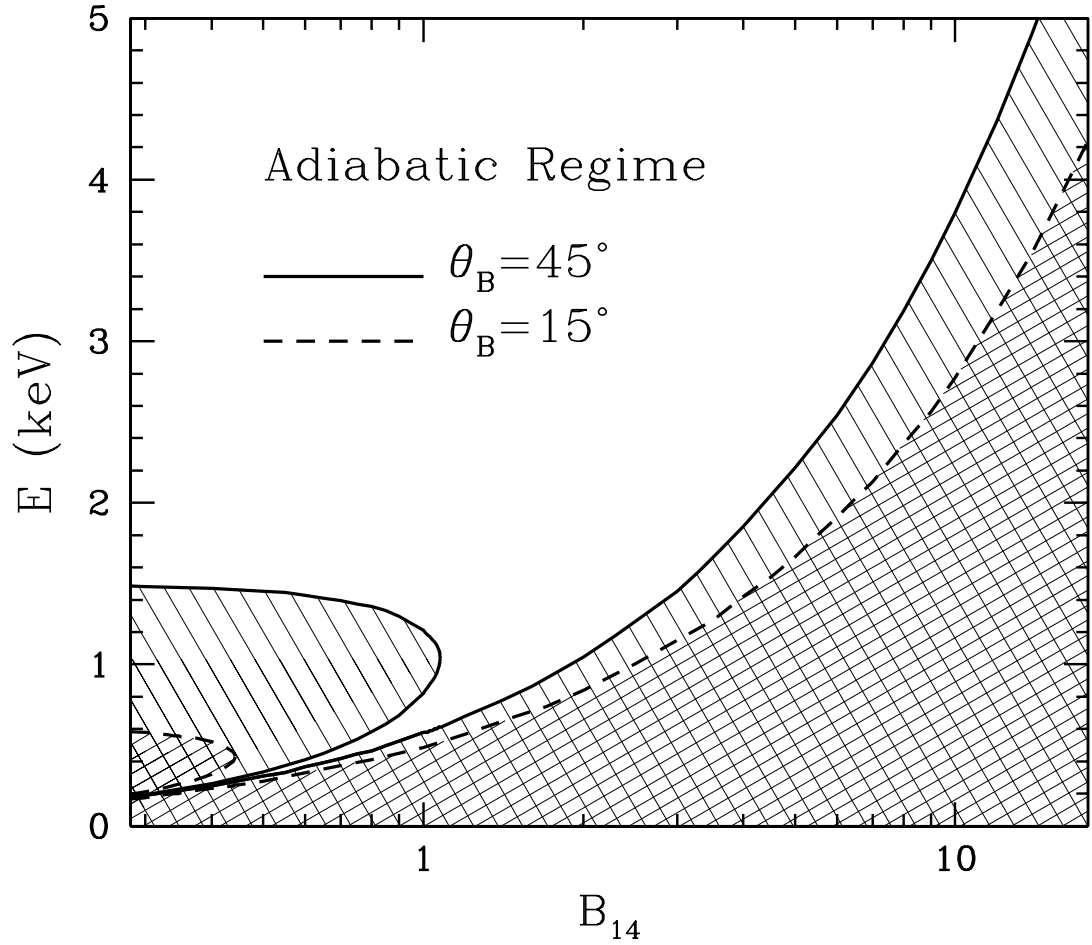}
\caption{The critical photon energy for adiabatic mode conversion
as a function of magnetic field strength. The unshaded region 
refers to the adiabatic regime in which the condition $E>E_{\rm ad}$ 
is satisfied [see eq.~(\ref{condition})].
The solid line is for $\theta_B=45^\circ$ and the dashed line for
$\theta_B=15^\circ$, both assuming $H_\rho=5$~cm.
\label{fig6}}
\end{figure*}

\begin{figure*}
\plotone{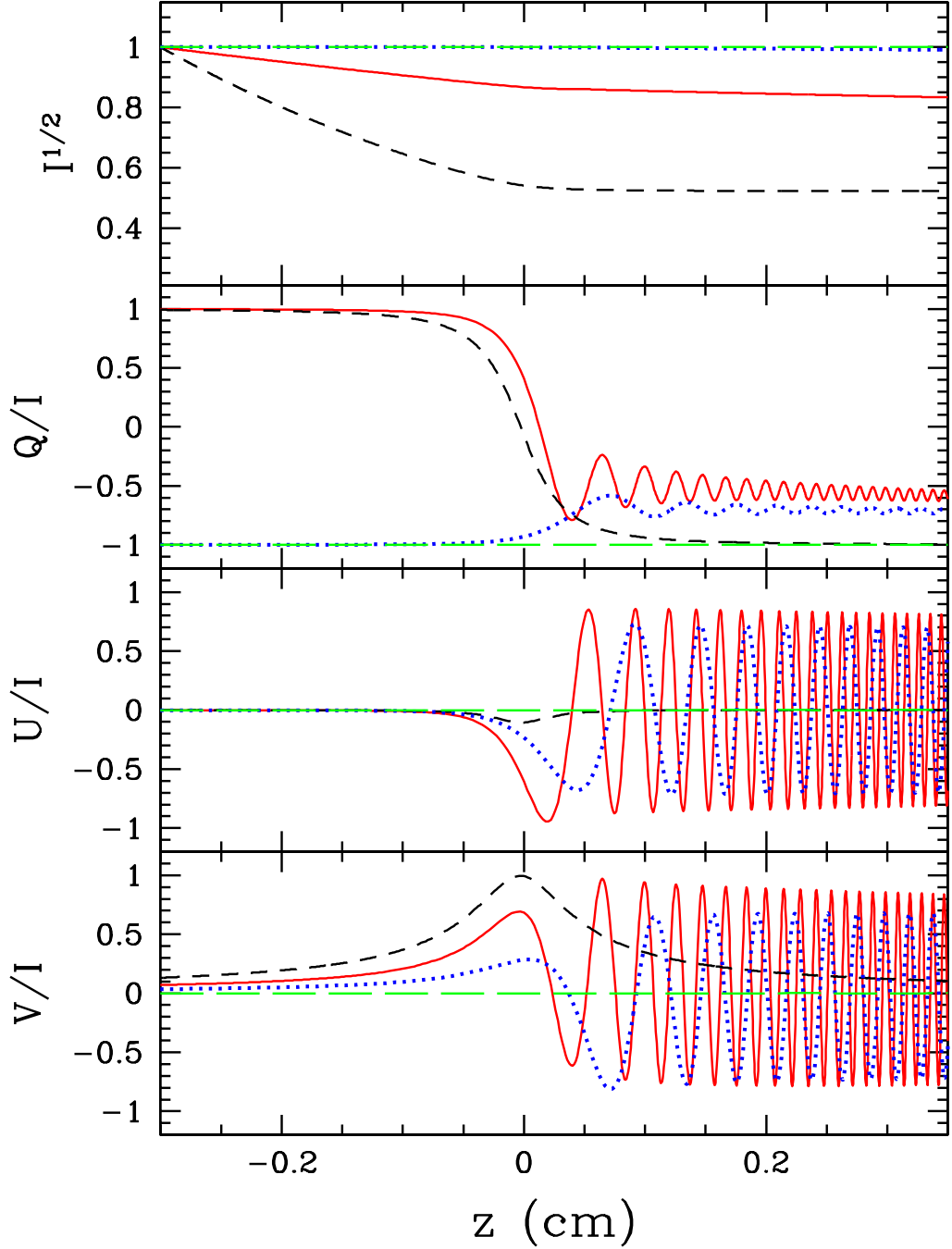}
\caption{Evolution of Stokes parameters across the vacuum resonance.
The parameters are $B_{14}=1$, $\theta_B=45^\circ$, and the 
density profile is $\rho=\rho_V \exp({-z/H_\rho})$ with $H_\rho=5$~cm.
At $z=-0.3$~cm, the radiation is in the plus-mode, with $I=1$.
The solid lines are for $E=1$~keV, the short-dashed lines for 
$E=3$~keV, the dotted lines for $E=0.5$~keV, and the long-dashed
lines for $E=0.1$~keV.
\label{fig7}}
\end{figure*}

\begin{figure*}
\plotone{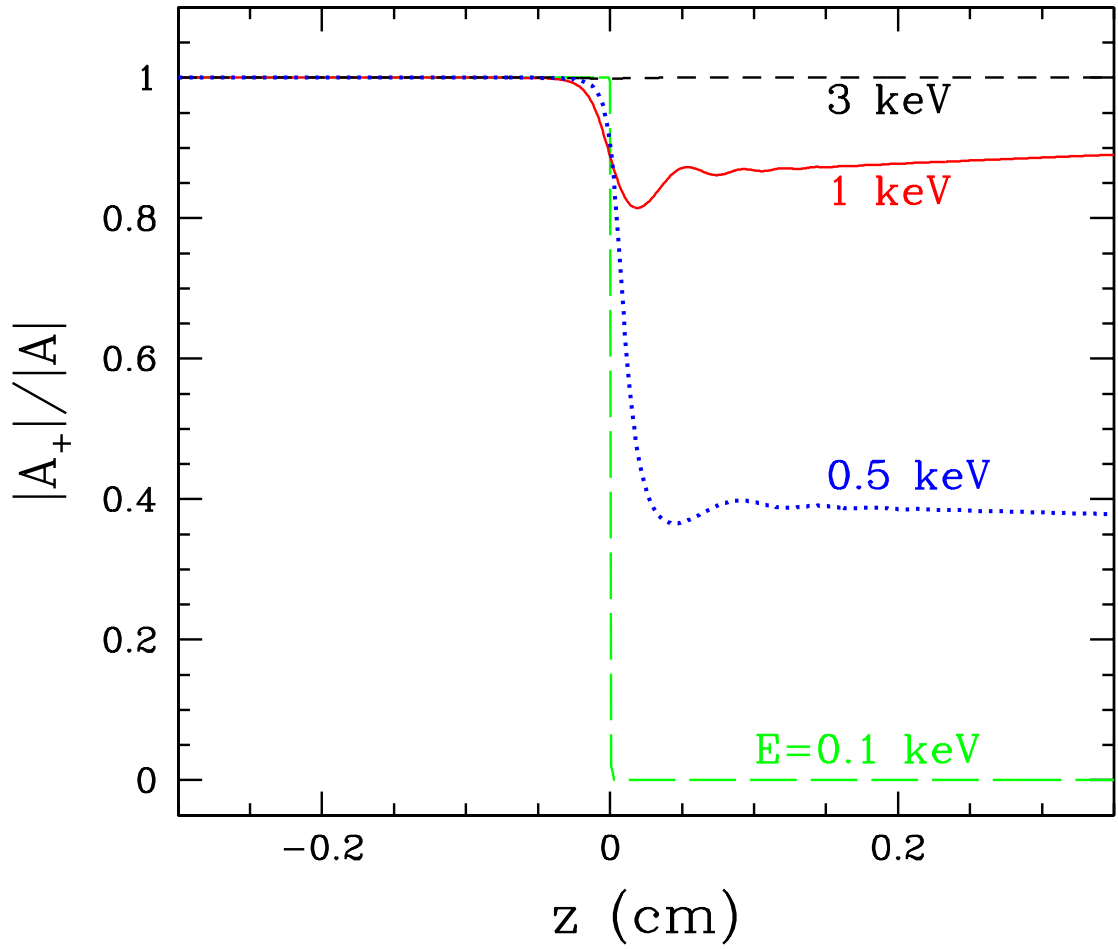}
\caption{Evolution of the ratio $|A_+|/|A|$ 
for the cases depicted in Fig.~7. Here $|A_\pm|$ is the amplitude of
the plus and minus-modes and $|A|=\sqrt{|A_+|^2+|A_-|^2}$.
\label{fig8}}
\end{figure*}

\begin{figure*}
\plotone{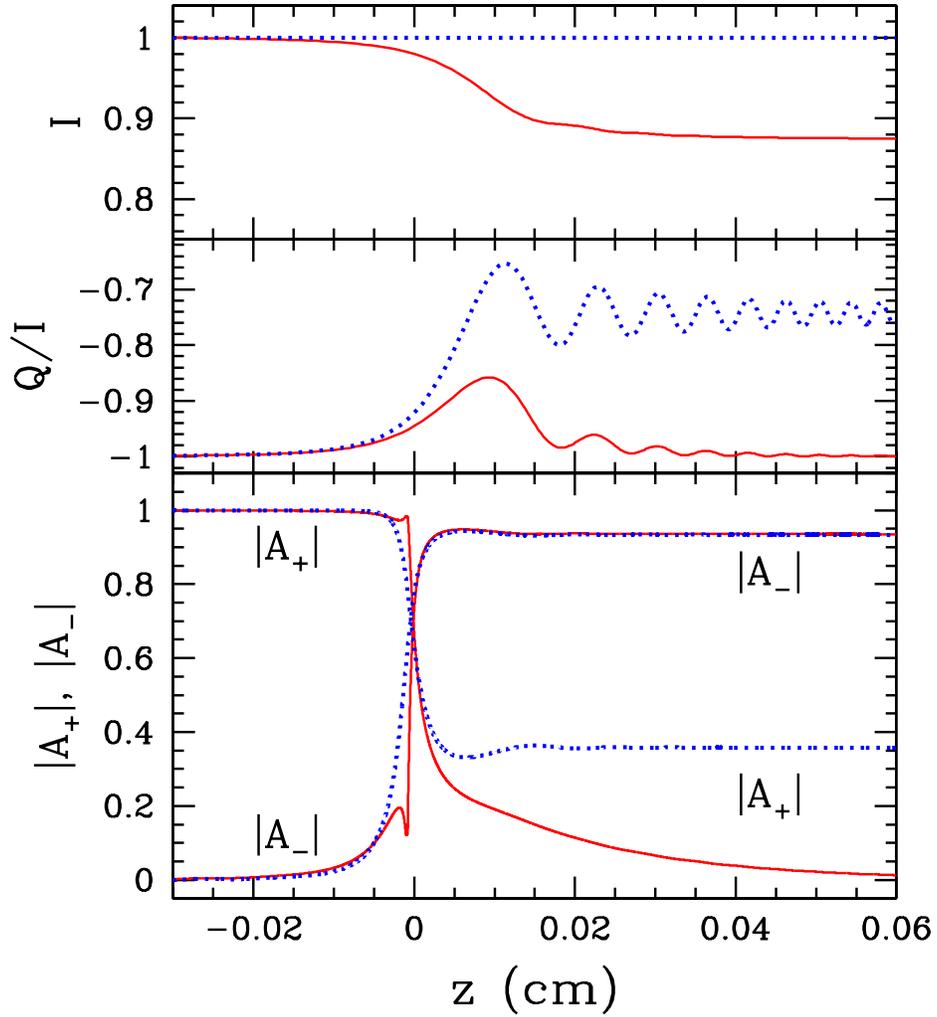}
\caption{Evolution of Stokes parameters $I$, $Q$ and mode amplitudes
$|A_+|,~|A_-|$ across vacuum resonance, for $E=1.7$~keV,
$B_{14}=5$, $\theta_B=45^\circ$, and $H_\rho=5$~cm.
At $z=-0.03$~cm, the radiation is in the plus-mode, with $A_+=1$.
The solid lines show the results 
including realistic damping terms in the dielectric
tensor, and the dotted lines show the results when the damping terms are
turned off. Note that for this case, $\beta=1.066\,i$ at $\rho=\rho_V$, 
so that the mode amplitudes are well defined only
away from the the resonance ($A_+$ and $A_-$ are calculated assuming
that $\theta_m$ is real and the two modes are completely orthogonal).
\label{fig9}}
\end{figure*}

\end{document}